\documentclass[twocol]{ametsocV5}

\usepackage{amsmath,amsfonts,amssymb,bm}
\usepackage{mathptmx}
\usepackage{newtxtext}
\usepackage{newtxmath}

\newcommand{\inner}[2]{\left< #1, #2 \right>} 

\renewcommand{\vec}[1]{\boldsymbol{#1}  } 

\renewcommand{\d}[2]{\frac{\mathrm{d} #1}{\mathrm{d} #2}} 
\newcommand{\dd}[2]{\frac{\mathrm{d}^2 #1}{\mathrm{d} #2^2}} 
\newcommand{\pd}[2]{\frac{\partial #1}{\partial #2}} 
\newcommand{\grad}[1]{\vec{\nabla} #1} 
\newcommand{\lap}{\nabla^2} 
\newcommand{\unit}[1]{\hat{\vec{#1}}}

\newcommand{\wh}[1]{\widehat{#1} }

\newcommand{\Dc}{{\mathcal{D}}}

\newcommand{\Qf}{\left(Q,R_1,R_2\right)}

\newcommand{\C}{\mathbb{C}}
\newcommand{\Pb}{\mathscr{P}}

\newcommand{\intz}{\int_{z_1}^{z_2}}

\title{On the Discrete Normal Modes of Quasigeostrophic Theory}

\authors{Houssam Yassin\correspondingauthor{Houssam Yassin, hyassin@princeton.edu}}

\extraauthor{Stephen M. Griffies}

\affiliation{Program in Atmospheric and Oceanic Sciences, Princeton University, Princeton, NJ, USA}

\extraaffil{NOAA/Geophysical Fluid Dynamics Laboratory and \\
Program in Atmospheric and Oceanic Sciences, Princeton University, Princeton, NJ, USA}

\abstract{The discrete baroclinic modes of quasigeostrophic theory are incomplete and the incompleteness manifests as a loss of information in the projection process. The incompleteness of the baroclinic modes is related to the presence of two previously unnoticed stationary step-wave solutions of the Rossby wave problem with flat boundaries. These step-waves are the limit of surface quasigeostrophic waves as boundary buoyancy gradients vanish. A complete normal mode basis for quasigeostrophic theory is obtained by considering the traditional Rossby wave problem with prescribed buoyancy gradients at the lower and upper boundaries. The presence of these boundary buoyancy gradients activates the previously inert boundary degrees of freedom. These Rossby waves have several novel properties such as the presence of multiple modes with no internal zeros, a finite number of modes with negative norms, and their vertical structures form a basis capable of representing any quasigeostrophic state with a differentiable series expansion. Using this complete basis, we are able to obtain a differentiable series expansion to the potential vorticity of Bretherton (with Dirac delta contributions). We also examine the quasigeostrophic vertical velocity modes and derive a complete basis for such modes as well. A natural application of these modes is the development of a weakly non-linear wave-interaction theory of geostrophic turbulence that takes topography into account.}
\begin{document}

\maketitle

%
%
%

%








\section{Introduction}

\subsection{Background}

The vertical decomposition of quasigeostrophic motion into normal modes plays an important role in bounded stratified geophysical fluids \citep[e.g.,][]{charney_geostrophic_1971,flierl_models_1978,fu_nonlinear_1980,wunsch_vertical_1997, chelton_geographical_1998, smith_scales_2001, tulloch_quasigeostrophic_2009, lapeyre_what_2009, Ferrari2010a,ferrari_distribution_2010, de_la_lama_vertical_2016, lacasce_prevalence_2017,brink_structure_2019}. Most prevalent are the traditional baroclinic modes \cite[e.g., section 6.5.2 in][]{Vallis2017} that are the vertical structures of Rossby waves in a quiescent ocean with no topography or boundary buoyancy gradients. In a landmark contribution, \cite{wunsch_vertical_1997} partitions the ocean's kinetic energy into the baroclinic modes and finds that the zeroth and first baroclinic modes dominate over most of the extratropical ocean. Additionally, \cite{wunsch_vertical_1997} concludes that the surface signal primarily reflects the first baroclinic mode and, therefore, the motion of the thermocline.

However, the use of  baroclinic modes has come under increasing scrutiny in recent years \citep{lapeyre_what_2009,roullet_properties_2012,scott_assessment_2012,smith_surface-aware_2012}. \cite{lapeyre_what_2009} observes that the vertical shear of the baroclinic modes vanishes at the boundaries, thus leading to the concomitant vanishing of the boundary buoyancy. Consequently, \cite{lapeyre_what_2009} proposes that the baroclinic modes cannot be complete\footnote{A collection of functions is said to be complete in some function space, $\mathcal{F}$, if this collection forms a basis of $\mathcal{F}$. Specifying the underlying function space, $\mathcal{F}$, turns out to be crucial, as we see in section \ref{S-phase}\ref{S-vertical-phase}.}  due to their inability to represent boundary buoyancy. To supplement the baroclinic modes, \cite{lapeyre_what_2009} includes a boundary-trapped exponential surface quasigeostrophic solution \cite[see][]{held_surface_1995} and suggests that the surface signal primarily reflects, not thermocline motion, but boundary-trapped surface quasigeostrophic dynamics \cite[see also][]{lapeyre_surface_2017}.

Appending additional functions to the collections of normal modes as in \cite{lapeyre_what_2009} or \cite{scott_assessment_2012} does not result in a set of normal modes since the appended functions are not  orthogonal to the original modes. It is only with \cite{smith_surface-aware_2012} that a set of normal modes capable of representing arbitrary surface buoyancy is derived.

Yet it is not clear how the normal modes of \cite{smith_surface-aware_2012} differ from the baroclinic modes or what these modes correspond to in linear theory. Indeed, \cite{rocha_galerkin_2015}, noting that the baroclinic series expansion of any sufficiently smooth function converges uniformly to the function itself,  argues that the incompleteness of the baroclinic modes has been ``overstated''. Moreover, \cite{de_la_lama_vertical_2016} and \cite{lacasce_prevalence_2017}, motivated by the observation that the leading empirical orthogonal function of \cite{wunsch_vertical_1997} vanishes near the ocean bottom, propose an alternate set of modes\textemdash the surface modes\textemdash that have a vanishing pressure at the bottom boundary.

We thus have a variety of proposed normal modes and it is not clear how their properties differ. Are the baroclinic modes actually incomplete? What about the surface modes? What does completeness mean in this context? 
The purpose of this paper is to  answer these questions. 

\subsection{Normal modes and eigenfunctions}

A normal mode is a linear motion in which all components of a system move coherently at a single frequency. Mathematically, a normal mode has the form
\begin{equation}\label{physical-mode}
	 \Phi_a( x,y,z) \, \mathrm{e}^{-\mathrm{i} \omega_a t}
\end{equation} 
where $\Phi_a$ describes the spatial structure of the mode and $\omega_a$ is its angular frequency. The function $\Phi_a$ is obtained by solving a differential eigenvalue problem and hence is an eigenfunction. The collection of all eigenfunctions forms a basis of some function space relevant to the problem. 

By an abuse of terminology, the spatial structure, $\Phi_a$, is often called a normal mode (e.g., the term ``Fourier mode'' is often used for $\mathrm{e}^{\mathrm{i} k \, x}$ where $k$ is a wavenumber). In linear theory, this misnomer is often benign as each $\Phi_a$ corresponds to a frequency $\omega_a$. For example, given some initial condition $\Psi(x,y,z)$, we decompose $\Psi$ as a sum of modes at $t=0$,
\begin{equation}\label{sum-of-eigen}
	\Psi(x,y,z) = \sum_{a}  c_a \, \Phi_a(x,y,z),
\end{equation}
where the $c_a$ are the Fourier coefficients, and the time evolution is then given by
\begin{equation}\label{sum-of-eigen-time}
	\sum_{a} c_a \, \Phi_a(x,y,z) \, \mathrm{e}^{-\mathrm{i} \omega_a t}.
\end{equation}

However, with non-linear dynamics, this abuse of terminology can be confusing. Given some spatial structure, $\Psi(x,y,z)$, in a fluid whose flow is non-linear, we can still exploit the basis properties of the eigenfunctions $\Phi_a$ to decompose $\Psi$ as in equation \eqref{sum-of-eigen}. Whereas in a linear fluid only wave motion of the form \eqref{physical-mode} is possible, a non-linear flow admits a larger collection of solutions (e.g., non-linear waves and coherent vortices) and so the linear wave solution \eqref{sum-of-eigen-time} no longer follows from the decomposition \eqref{sum-of-eigen}. 

For this reason, we call the linear solution \eqref{physical-mode} a \emph{physical} normal mode to distinguish it from the spatial structure $\Phi_a$, which is only an eigenfunction. Otherwise, we will use the terms ``normal mode'' and ``eigenfunction'' interchangeably to refer to the spatial structure $\Phi_a$, as is prevalent in the literature.

Our strategy here is then the following. We find the \emph{physical} normal modes [of the form \eqref{physical-mode}] to various Rossby wave problems  and examine the basis properties of their constituent eigenfunctions $\Phi_a$. Our goal is to find a collection of eigenfunctions (i.e., ``normal modes'' in the prevalent terminology) capable of representing every possible quasigeostrophic state.

\subsection{Contents of this article}

This article constitutes an examination of all collections of discrete (i.e., non-continuum\footnote{Continuum modes appear once a sheared mean-flow is present, e.g., \cite{drazin_rossby_1982}, \cite{balmforth_normal_1994,balmforth_singular_1995}, and \cite{brink_structure_2019}. }) quasigeostrophic normal modes. We include the baroclinic modes, the surface modes of \cite{de_la_lama_vertical_2016} and \cite{lacasce_prevalence_2017}, the surface-aware mode of \cite{smith_surface-aware_2012}, as well as various generalizations. To study the completeness of a set of normal modes, we must first define the underlying space in question. From general considerations, we introduce in section \ref{S-phase} the quasigeostrophic phase space, defined as the space of all possible quasigeostrophic states. Subsequently, in section \ref{S-linear} we use the general theory of differential eigenvalue problems with eigenvalue dependent boundary conditions, as developed in \cite{yassin_normal_nodate}, to study Rossby waves in an ocean with prescribed boundary buoyancy gradients (e.g., topography, see section \ref{S-phase}\ref{SS-pv}). Intriguingly, in an ocean with no topography, we find that, in addition to the usual baroclinic modes, there are two additional stationary step-mode solutions that have not been noted before. The stationary step-modes are the limits of boundary-trapped surface quasigeostrophic waves as the boundary buoyancy gradient vanishes. 

Our study of Rossby waves then leads us  examine  all possible discrete collections of normal modes in section \ref{S-expansions}.
As shown in this section,  the baroclinic modes are incomplete, as argued by \cite{lapeyre_what_2009}, and we point out that the incompleteness leads to a \emph{loss} of information after projecting a function onto the baroclinic modes.
In contrast, modes such as those suggested by \cite{smith_surface-aware_2012} are complete in the quasigeostrophic phase space so that projecting a function onto such modes provides an \emph{equivalent} representation of the function. 

We offer discussion of our analysis in Section \ref{S-discussion} and conclusions in Section \ref{S-conclusion}. Appendix A summarizes the key mathematical results pertaining to eigenvalue 
problems where the eigenvalue appears in the boundary conditions.  Appendix B then summarizes the polarization relations as well as the vertical velocity eigenvalue problem.

\section{Mathematics of the quasigeostrophic phase space}\label{S-phase}

\subsection{The potential vorticity}\label{SS-pv}


Consider a three-dimensional region $\Dc$ of the form 
\begin{equation}\label{region}
	\Dc = \Dc_0 \times \left[z_1,z_2\right].
\end{equation}
The area of the lower and upper boundaries is denoted by $\Dc_0$ and is a rectangle of area $A$ while $z_1$ (lower boundary) and $z_2$ (upper boundary) are constants. The horizontal boundaries are either rigid or periodic. 

The state of a quasigeostrophic fluid in $\Dc$ is determined by a charge-like quantity known as the quasigeostrophic potential vorticity \citep{hoskins_use_1985,schneider_boundary_2003}. If the potential vorticity is distributed throughout the three-dimensional region $\Dc$, we are concerned with the volume potential vorticity density, $Q$, with $Q$  related to the geostrophic streamfunction $\psi$ by [e.g., section 5.4 of \cite{Vallis2017}]
\begin{equation}\label{q-psi}
	Q = f + \lap \psi + \pd{}{z} \left(\frac{f_0^2}{N^2} \pd{\psi}{z} \right).
\end{equation}
Here, the latitude dependent Coriolis parameter is 
\begin{equation}
	f = f_0 + \beta\,y,
\end{equation}
$N(z)$ is the prescribed background buoyancy frequency,  $\nabla^{2}$ is the horizontal Laplacian operator, and 
\begin{equation}
\vec u = \unit z \times \grad \psi
\end{equation}
is the horizontal geostrophic velocity, $\vec u = (u,v)$. 

Additionally, the potential vorticity may be distributed over a two-dimensional region, say the lower and upper boundaries $\Dc_0$, to obtain surface potential vorticity densities $R_1$ and $R_2$.  The surface potential vorticity densities are related to the streamfunction by
\begin{equation}\label{r-psi}
	R_j = (-1)^{j+1} \left[ g_j + \left( \frac{f_0^2}{N^2} \, \pd{\psi}{z}\right)\Bigg |_{z=z_j}  \right]
\end{equation}
where $g_j$ is an imposed surface potential vorticity density at the lower or upper boundary and $j=1,2$. The density $g_j$ corresponds to a prescribed buoyancy 
\begin{equation}
 b_j =	\frac{N^2}{f_0} g_j
\end{equation}
at the $j$th boundary [see equation \eqref{buoyancy-boundary-evolution}]. Alternatively, $g_j$ may be thought of as an infinitesimal topography through
\begin{equation}\label{g_j_topography}
	g_j = f_0 h_j
\end{equation}
where $h_j$ represents infinitesimal topography at the $j$th boundary. Whereas $Q$ has dimensions of inverse time, $R_{j}$ has dimensions of length per time.


\subsection{Defining the quasigeostrophic phase space}

We define the quasigeostrophic phase space to be the space of all possible quasigeostrophic states, with a quasigeostrophic state  determined by the potential vorticity densities, $Q, R_1$, and $R_2$. Note that the volume potential vorticity density, $Q$, is defined throughout the whole fluid region $\Dc$, so that $Q = Q(x,y,z,t)$. In contrast, the surface potential vorticity densities, $R_1$ and $R_2$, are only defined on the two-dimensional lower and upper boundary surfaces, $\Dc_0$, so that $R_j = R_j(x,y,t)$.

It is useful to restate the previous paragraph with some added mathematical precision. For that purpose, let $L^2[\Dc]$ be the space of square-integrable functions\footnote{The definition of $L^2[\Dc]$ is   more subtle than presented here. Namely, elements of $L^2[\Dc]$ are not functions, but rather equivalence classes of functions leading to the unintuitive properties seen in this section. See \cite{yassin_normal_nodate} and citations within for more details.} in the fluid volume $\Dc$, and let $L^2[\Dc_0]$ be the space of square-integrable functions on the boundary area $\Dc_0$. Elements of $L^2[\Dc]$ are functions of three spatial coordinates whereas elements of $L^2[\Dc_0]$ are functions of two spatial coordinates. Hence, $Q \in L^2[\Dc]$ and $R_1,R_2 \in L^2[\Dc_0]$.

Define the space $\Pb$ by
\begin{equation}\label{phase-space}
	\Pb = L^2[\Dc]\oplus L^2[\Dc_0] \oplus L^2[\Dc_0],
\end{equation} 
where $\oplus$ is the direct sum. Equation \eqref{phase-space} states that any element of $\Pb$ is a tuple  $(Q,R_1,R_2)$ of three functions, where $Q=Q(x,y,z,t)$ is a function on the volume $\Dc$ and hence element of $L^2[\Dc]$, while the functions $R_j=R_j(x,y,t)$, for $j=1,2$, are functions on the area $\Dc_0$ and hence are elements of $L^2[\Dc_0]$. We conclude that $(Q,R_1,R_2) \in \Pb$ and that $\Pb$ is the space of all possible quasigeostrophic states. We thus call $\Pb$ the quasigeostrophic phase space.

\subsection{The phase space in terms of the streamfunction}


Given an element $\Qf \in \Pb$, we can reconstruct a continuous function $\psi$ that contains the same dynamical information as $\Qf$. By inverting the problem
\begin{align}\label{int-pv-psi}
\begin{split}
	Q-f =  \lap \psi_{\textrm{int}}  + \pd{}{z} \left( \frac{f_0}{N^2} \pd{ \psi_{\textrm{int}}}{z} \right) \quad &\text{for } z \in (z_1,z_2) \\
	R_1 - g_1 = \frac{f_0^2}{N^2}\, \pd{ \psi_{\textrm{low}} }{z} \quad &\text{for } z=z_1\\
	R_2 + g_2  = -\frac{f_0^2}{N^2} \, \pd{ \psi_{\textrm{upp}}}{z} \quad &\text{for } z=z_2
\end{split}
\end{align}
we obtain a function $\psi(x,y,z)$ that is unique up to a gauge transformation \cite[see][]{schneider_boundary_2003}. Conversely, given a function $\psi(x,y,z)$, we can differentiate $\psi$ as in equations \eqref{int-pv-psi} to obtain $\Qf\in\Pb$. Thus, we can also consider the quasigeostrophic phase space $\Pb$ to be the space of all possible streamfunctions $\psi$.

Equations \eqref{int-pv-psi} motivate the definition of the relative potential vorticity densities, $q=Q-f$ and $r_j= R_j - (-1)^{j+1}\, g_j$, which are the portions of the potential vorticity providing a source for a streamfunction. Explicitly, the relative potential vorticity densities are 
\begin{subequations}
\begin{alignat}{2}
	q &= \lap \psi  + \pd{}{z} \left( \frac{f^2_0}{N^2} \pd{ \psi}{z}\right) 
	\quad &&\text{for } z \in (z_1,z_2)  
	\\
	r_1 &= \frac{f_0^2}{N^2} \pd{\psi}{z} 
	 \quad &&\text{for } z=z_1 
	\\
	 r_2 &= -\frac{f_0^2}{N^2} \pd{\psi }{z} 
	 \quad &&\text{for } z=z_2.
\end{alignat}
\end{subequations}

\subsection{The vertical structure phase space}\label{S-vertical-phase}

Since the fluid region, $\Dc$, is separable, we can expand the potential vorticity density distribution, $(q,r_1,r_2)$, and  the streamfunction $\psi$ in terms of the eigenfunctions, $e_{\vec k}$, of the horizontal Laplacian. For a horizontal domain $\Dc_0$, the eigenfunction $ e_{\vec k}(\vec x)$ satisfies
\begin{equation}
	- \lap e_{\vec k} = k^2\, e_{\vec k}.
\end{equation}
where $\vec x=(x,y)$ is the horizontal position vector, $\vec k=(k_x,k_y)$ is the horizontal wavevector, and $k=|\vec k|$ is the horizontal wavenumber. For example, in a horizontally periodic domain the eigenfunctions $e_{\vec k}(\vec x)$ are proportional to complex exponentials, $\mathrm{e}^{\mathrm{i}\vec k \cdot \vec x}$. 

Projecting the relative potential vorticity density distribution, $(q,r_1,r_2)$, onto the horizontal eigenfunctions, $e_{\vec k}$, yields 
\begin{subequations}
\begin{alignat}{2}
	q(\vec x,z,t) &= \sum_{\vec k} q_{\vec k}(z,t) \, e_{\vec k}(\vec x), \quad &&\text{for } z\in(z_1,z_2)\\
	r_j(\vec x,t) &= \sum_{\vec k} r_{j \vec k}(t) \, e_{\vec k}(\vec x) \quad &&\text{for } j=1,2.
\end{alignat}
\end{subequations}
Thus the Fourier coefficients of $(q,r_1,r_2)$ are $(q_{\vec k}, r_{1\vec k}, r_{2\vec k})$ where $q_{\vec k}$ is a function of $z$ and $r_{1 \vec k}$ and $r_{2\vec k}$ are independent of $z$. Hence, $q_{\vec k}$ is an element of $L^2[(z_1,z_2)]$ whereas  $r_{1 \vec k}$ and $ r_{2 \vec k}$ are elements of the space of complex numbers\footnote{Since all physical fields must be real, only a single degree of freedom is gained from $\C$. Furthermore, when complex notation is used (e.g., complex exponentials for the horizontal eigenfunctions $e_{\vec k}$) it is only the real part of the fields that is physical. } , $\C$.

We conclude that the vertical structure of the potential vorticity, given by $(q_{\vec k}, r_{1\vec k}, r_{2\vec k})$, is an element of
\begin{equation}\label{vertical-phase-space}
	\wh \Pb = L^2[(z_1,z_2)] \oplus \C \oplus \C,
\end{equation}
so that the vertical structures of the potential vorticity distribution are determined by a function, $q_{\vec k}$, in $L^2[(z_1,z_2)]$ and two $z$-independent elements, $r_{1 \vec k}$ and $r_{2  \vec k}$, of $\C$.  Similarly, the streamfunction can be represented as 
\begin{equation}
	\psi(\vec x,z,t) = \sum_{\vec k} \psi_{\vec k} (z,t) \, e_{\vec k} (\vec x),
\end{equation}
where $\psi_{\vec k}$ and $(q_{\vec k}, r_{1\vec k}, r_{2\vec k})$ are related by
\begin{subequations}\label{pv-hor-transform}
\begin{align}
	q_{\vec k} = -k^2 \, \psi_{\vec k} + \pd{}{z} \left(\frac{f_0^2}{N^2} \pd{\psi_{\vec k}}{z}\right) \\
	r_{j \vec k} = (-1)^{j+1} \left(\frac{f_0^2}{N^2} \pd{\psi_{\vec k}}{z} \right)\Bigg |_{z=z_j}.
\end{align}
\end{subequations}

As before, knowledge of the vertical structure of the streamfunction, $\psi_{\vec k}(z)$, is equivalent to knowing the vertical structure of the potential vorticity distribution, $(q_{\vec k}, r_{1\vec k}, r_{2\vec k})$. Thus $\wh \Pb$ is also the space of all possible streamfunction vertical structures.

That $\psi_{\vec k}$ belongs to $\wh \Pb$ and not $L^2[(z_1,z_2)]$ underlies much of the confusion over baroclinic modes. Assertions of completeness, based on Sturm-Liouville theory, assume that $\psi$ is an element of $L^2[(z_1,z_2)]$. However, as we have shown, that is an incorrect assumption. That $\psi$ belongs to $\wh \Pb$ will have consequences for the convergence and differentiability of normal mode expansions, as discussed in section \ref{S-expansions}. In the context of quasigeostrophic theory, the space $\wh \Pb$ first appeared in \cite{smith_surface-aware_2012}. More generally, $\wh \Pb$  appears in the presence of  non-trivial boundary dynamics \citep{yassin_normal_nodate}.

We call $\wh \Pb$ the vertical structure phase space, and for convenience we denote $L^2[(z_1,z_2)]$ by $L^2$ for the remainder of the article. The vertical structure phase space $\wh \Pb$ is then written as the direct sum 
\begin{equation}
	\wh \Pb = L^2 \oplus \C^2.
\end{equation}

\subsection{Representing the energy and potential enstrophy}
\label{subsection:decomposing-energy-enstrophy}

We find it convenient to represent several quadratic quantities in terms of the eigenfunctions of the horizontal Laplacian, $e_{\vec k}(\vec x)$. The energy per unit mass in the volume $\Dc$ is given by
\begin{align}
	E = \frac{1}{V}\int_{\Dc} \left[ \left |\grad \psi \right |^2 + \frac{f_0^2}{N^2} \left |\pd{\psi}{z} \right |^2 \right] \mathrm{d}A \, \mathrm{d}z = \sum_{\vec k} E_{\vec k},
\end{align}
where the horizontal energy mode is given by the vertical integral 
\begin{equation}
	E_{\vec k} =  \frac{1}{H}\intz \left[ k^2 \left| \psi_{\vec k} \right|^2 + \frac{f_0^2}{N^2}  \left |\pd{\psi_{\vec k}}{z} \right |^2 \right] \mathrm{d}z,
\end{equation}
with $V=A\,H$ the domain volume and $H =z_2-z_1$ the domain depth.

Similarly, for the relative volume potential enstrophy density, $Z$, we have
\begin{align}
	Z &= \frac{1}{V} \int_{\Dc} |q|^2  \mathrm{d}A \, \mathrm{d}z = \sum_{\vec k} Z_{\vec k}
\end{align}
where 
\begin{equation}
	Z_{\vec k} = \frac{1}{H} \intz  \left|q_{\vec k} \right|^2 \mathrm{d}z.
\end{equation}
Finally, analogous to $Z$, we have the relative surface potential enstrophy densities, $Y_j$, on the area $\Dc_0$
\begin{equation}\label{surface-enstrophy}
	Y_j = \frac{1}{A} \int_{\Dc_0} \left| r_j \right|^2 \mathrm{d}A = \sum_{\vec k} Y_{j\vec k}
\end{equation}
where 
\begin{equation}\label{surface-enstrophy-rjk}
	Y_{j\vec k} =  \left | r_{j\vec k} \right|^2.
\end{equation}

\section{Rossby waves in a quiescent ocean}\label{S-linear}

\begin{figure}
	\noindent \includegraphics[width=\columnwidth]{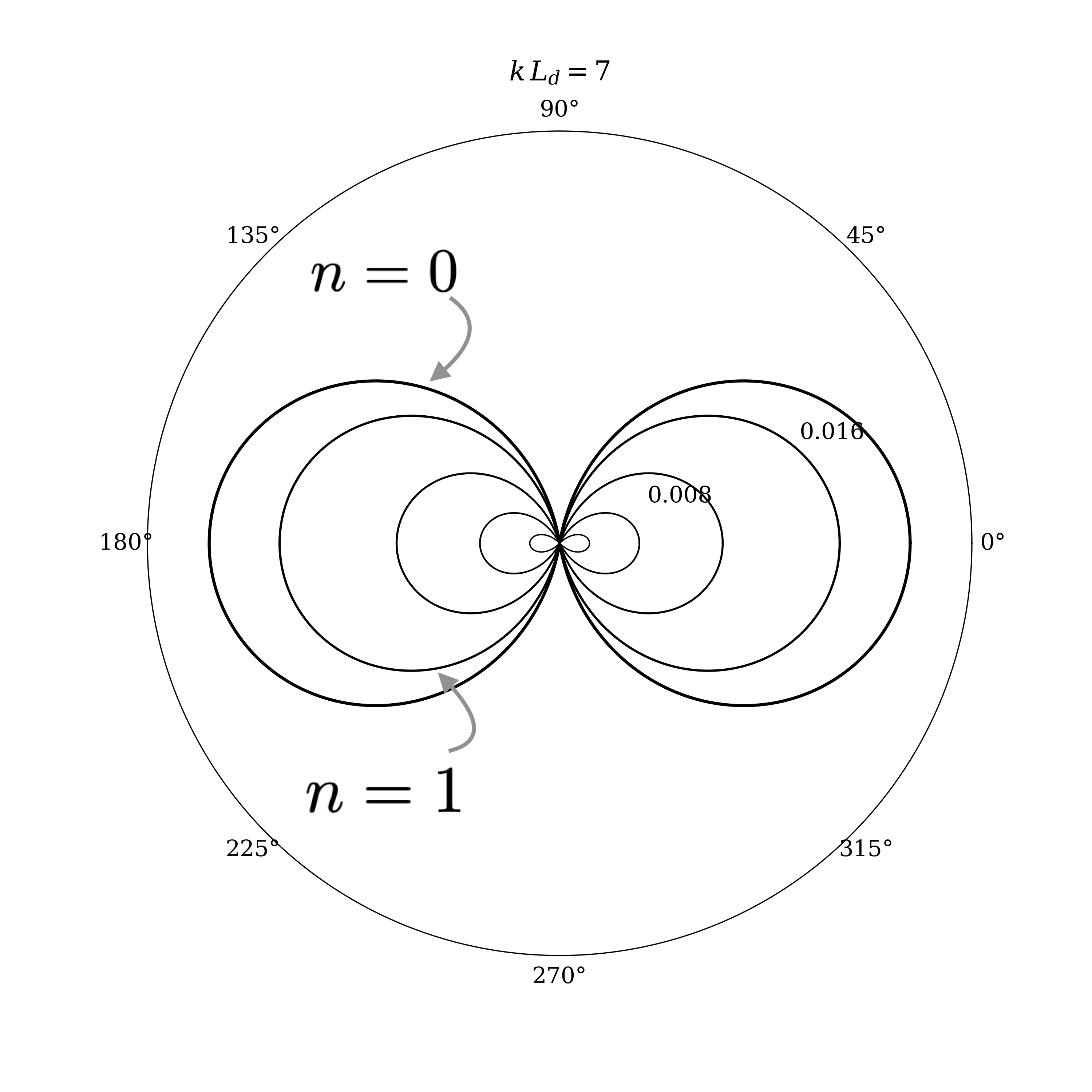}
	\caption{Polar plots of the absolute value of the non-dimensional angular frequency $|\omega_n|/(\beta L_d)$ of the first five modes of the traditional eigenvalue problem (section \ref{S-linear}\ref{SS-trad}) as a function of the wave propagation direction, $\vec k/|\vec k|$, for constant stratification. The outer most ellipse, with the largest absolute angular frequency, represents the angular frequency of the barotropic ($n=0$) mode. The higher modes have smaller absolute frequencies and are thus concentric and within the barotropic angular frequency curve. Since the absolute value of the angular frequency of the barotropic mode becomes infinitely large at small horizontal wavenumbers $k$, we have chosen a large wavenumber $k$, given by $k\,L_d = 7$, so that the angular frequency of the first five modes can be plotted in the same figure. We have chosen $f_0 = 10^{-4}~\textrm{s}^{-1},  \beta = 10^{-11}~\textrm{m}^{-1}~\textrm{s}^{-1}, N_0 = 10^{-2}~\textrm{s}^{-1}$ and $H = 1~\textrm{km}$ leading to a deformation radius $L_d = N_0 \, H/f_0= 100~\textrm{km}$. Numerical solutions to all eigenvalue problems in this paper are obtained using \texttt{Dedalus} \citep{burns_dedalus_2020}.}
	\label{F-baroclinic_angle}
\end{figure}

In this section, we study Rossby waves in an otherwise quiescent ocean; in other words, we examine the \emph{physical} normal modes of a quiescent ocean. The linear equations of motion are
\begin{subequations}\label{linear-time-evolution}
\begin{align}
	\label{linear-q-equation}
	\pd{q}{t} + \beta \, v = 0 \quad &\textrm{for } z\in(z_1,z_2)\\
	\label{linear-r-equation}
	\pd{r_{j}}{t} + \vec u  \cdot \grad\left[(-1)^{j+1} \, g_j\right]  = 0 \quad &\textrm{for } z=z_j.
\end{align}
\end{subequations}
We assume that the prescribed surface potential vorticity densities at the lower and upper boundaries, $g_1$ and $g_2$, are linear, which ensures the resulting eigenvalue problem is separable. Moreover, as the ocean is quiescent, $g_1$ and $g_2$ must refer to topographic slopes, as in equation \eqref{g_j_topography}.


The importance of the linear problem \eqref{linear-time-evolution} is that it provides all possible discrete Rossby wave normal modes in a quasigeostrophic flow. Substituting a wave ansatz of the form [compare with equation \eqref{physical-mode} for \emph{physical} normal modes]
\begin{equation}
	\psi(\vec x, z,t) = \hat \psi(z) \, e_{\vec k}(\vec x) \, \mathrm{e}^{-\mathrm{i}\omega t}
\end{equation}
into the  linear problem \eqref{linear-time-evolution} renders
\begin{equation}\label{q-eigen}
	\left(-\mathrm{i}\, \omega\right) \left[ -k^2 \, \hat \psi + \d{}{z}\left(\frac{f_0^2}{N^2} \d{\hat \psi}{z} \right) \right] + \mathrm{i} \, k_x \, \beta \, \hat \psi = 0 
\end{equation}
for $z\in(z_1,z_2)$, and 
\begin{equation}
	\label{r-eigen}
	\left(-\mathrm{i} \, \omega \right) \left( \frac{f_0^2}{N^2} \d{\hat \psi}{z} \right) + \mathrm{i} \, \unit z \cdot \left( \vec k \times \grad g_j \right) \hat \psi = 0 
\end{equation}
for $z=z_1,z_2$.

\subsection{Traditional Rossby wave problem}\label{SS-trad}

We first examine the traditional case of linear fluctuations to a quiescent ocean with isentropic lower and upper boundaries i.e., with no topography. Setting $\grad g_1=\grad g_2 = 0$ in the eigenvalue problem \eqref{q-eigen}\textendash\eqref{r-eigen} gives
\begin{subequations}\label{trad-rossby}
\begin{align}
	\label{trad-rossby-interior}
		\omega \left[ -k^2 \,  F + \d{}{z} \left(\frac{f_0^2}{N^2} \d{ F}{z}\right) \right] - \beta \, k_x \,  F = 0 \\
	\label{trad-rossby-boundary}
		\omega \left( \frac{f_0^2}{N^2} \d{F}{z}\right)\Bigg |_{z=z_j} = 0,
\end{align} 
\end{subequations}
where $\hat \psi(z) = \hat \psi_0 \, F(z)$ and $F$ is a non-dimensional function.
There are two cases to consider depending on whether $\omega$ vanishes.

\subsubsection{Traditional baroclinic modes} 

Assuming $\omega \neq 0$ in the eigenvalue problem \eqref{trad-rossby} renders a Sturm-Liouville eigenvalue problem in $L^2$
\begin{subequations}\label{trad-SL}
\begin{align}
	\label{trad-SL-interior}
	- \d{}{z}\left(\frac{f_0^2}{N^2}\d{ F}{z}\right) = \lambda \,  F \quad &\textrm{for } z\in (z_1,z_2) \\
	\label{trad-SL-boundary}
	\frac{f_0^2}{N^2} \d{ F}{z} = 0 \quad &\textrm{for } z=z_1,z_2,
\end{align}
\end{subequations}
where the eigenvalue, $\lambda$, is given by 
\begin{equation}\label{eigenvalue}
	\lambda = -k^2 - \frac{\beta \, k_x}{\omega}.
\end{equation}
See figure \ref{F-baroclinic_angle} for an illustration of the dependence of $|\omega|$ on the wavevector $\vec k$.

From Sturm-Liouville theory \cite[e.g.,][]{brown_fourier_1993}, the eigenvalue problem \eqref{trad-SL} has infinitely many eigenfunctions,  $F_0, \, F_1, \, F_2, \dots$ with distinct and ordered eigenvalues, $\lambda_n$, satisfying
\begin{equation}
	0 = \lambda_0 < \lambda_1 < \cdots \rightarrow \infty.
\end{equation}
The $n$th mode, $F_n$, has $n$ internal zeros in the interval $(z_1,z_2)$. 
The eigenfunctions are orthonormal with respect to
the inner product, $\left[\cdot ,\cdot\right]$, given by the vertical integral
\begin{equation}\label{trad-inner}
	\left[F , G\right] = \frac{1}{H} \intz F \, G \, \mathrm{d}z,
\end{equation}
with orthonormality meaning that  
\begin{equation}\label{trad-ortho}
	\delta_{mn} = \left[F_m,F_n\right]
\end{equation}
where $\delta_{mn}$ is the Kronecker delta. A powerful and commonly used result of Sturm-Liouville theory is that the set $\{F_n\}_{n=0}^\infty$ forms an orthonormal basis of $L^2$.

\subsubsection{Stationary step-modes}\label{SSS-step}

There are two additional solutions to the Rossby wave eigenvalue problem \eqref{trad-rossby} not previously noted in the literature. If $\omega = 0$ then the eigenvalue problem \eqref{trad-rossby} becomes
\begin{subequations}\label{trad-step}
\begin{align}
	\label{trad-step-interior}
		\beta \, k_x \, F = 0 \quad &\textrm{for } z\in(z_1,z_2) \\
	\label{trad-step-boundary}
		0 = 0 \quad &\textrm{for } z=z_1,z_2.
\end{align}
\end{subequations}
Consequently, if $k_x \neq 0$, then  $ F (z) =0$ for $z \in (z_1,z_2)$. That is, $F$ must vanish in the interior of the interval. However, since $\omega=0$ in \eqref{trad-rossby-boundary}, we obtain tautological boundary conditions \eqref{trad-step-boundary}. As a result, $F$ can take arbitrary values at the lower and upper boundaries. Thus two solutions are 
\begin{equation}\label{step-mode}
	F^\textrm{step}_{j}(z) = 
	\begin{cases}
		1  \quad \text{for } z=z_j\\
		0  \quad \text{otherwise.}
	\end{cases}
\end{equation}
The two step-mode solutions \eqref{step-mode} are independent of the traditional baroclinic modes, $F_{n}(z)$. An expansion of the step-mode $F^\textrm{step}_j$ in terms of the baroclinic modes will fail and produce a series that is identically zero.

The two stationary step-modes, $F^\textrm{step}_1$ and $F^\textrm{step}_2$, correspond to the two inert degrees of freedom in the eigenvalue problem \eqref{trad-rossby}. These two solutions are neglected in the traditional eigenvalue problem \eqref{trad-SL} through the assumption that $\omega \neq 0$. Although dynamically trivial, we will see that these two step-waves are obtained as limits of boundary-trapped modes as the boundary buoyancy gradients $N^2\,\grad g_j/f_0$ become small.

\subsubsection{The general solution}

For a wavevector $\vec k$ with $k_x \neq 0$, the vertical structure of the streamfunction must be of the form 
\begin{equation}\label{trad-vertical}
	\Psi(z) + \sum_{j=1}^2 \Psi_j \, F_j^\textrm{step}(z) = \psi_{\vec k} (z,t=0) 
\end{equation}
where $\Psi(z)$ is a twice differentiable function satisfying $\mathrm{d}\Psi(z_j)/\mathrm{d}z = 0$ for $j=1,2$ and $\Psi_1,\Psi_2$ are arbitrary constants. We can represent $\Psi$ according to the expansion
\begin{equation}\label{trad-expansion}
	\Psi = \sum_{n=0}^\infty \left[\Psi,F_n\right] F_n
\end{equation}
and so the time-evolution is 
\begin{equation}
	\psi_{\vec k} (z, t)= \sum_{n=0}^\infty  \left[\Psi,F_n\right] F_n \, \mathrm{e}^{-\mathrm{i}\omega_n t} + \sum_{j=1}^2 \Psi_j \, F^\textrm{step}_j.
\label{eq:linear-traditional-time-evolution}
\end{equation}
It is this time-evolution expression, which is valid only in linear theory for a quiescent ocean, that gives the baroclinic modes a clear physical meaning. More precisely, equation \eqref{eq:linear-traditional-time-evolution} states that the vertical structure $\Psi(z)$ disperses into its constituent Rossby waves with vertical structures $F_n$. Outside the linear theory of this section, baroclinic modes do not have a physical interpretation, although they remain a mathematical basis for $L^2$.

\subsection{The Rhines problem}\label{SS-rhines-bottom}

We now examine the case with a sloping lower boundary, $\grad g_1\neq 0$, and an isentropic upper boundary, $\grad g_2 =0$. The special case of a meridional bottom slope and constant stratification was first investigated by \cite{rhines_edge_1970}. Subsequently, \cite{charney_oceanic_1981} extended the analysis to realistic stratification and \cite{straub_dispersive_1994} examined the dependence of the waves on the propagation direction.  \cite{yassin_normal_nodate} applies the mathematical theory of eigenvalue problems with $\lambda$-dependent boundary conditions and obtains various completeness and expansion results as well as a qualitative theory for the streamfunction modes. Below, we generalize these results, study the two limiting boundary conditions, and consider the corresponding vertical velocity modes.

\subsubsection{The eigenvalue problem}

Let $\hat \psi(z) = \hat \psi_0 \, G(z)$ where $G$ is a non-dimensional function. We then manipulate the eigenvalue problem \eqref{q-eigen}\textendash\eqref{r-eigen} to obtain (assuming $\omega \neq 0$)
\begin{subequations}\label{Rhines-eigen_bottom}
\begin{alignat}{2}	
		- \d{}{z} \left(\frac{f_0^2}{N^2} \d{G}{z}\right) &= \lambda \, G \quad &&\textrm{for } z\in(z_1,z_2) \\
			-k^2 G - \gamma_1^{-1} \left( \frac{f_0^2}{N^2}\d{G}{z} \right) &= \lambda \, G \quad &&\textrm{for } z=z_1,\\
			\d{G}{z} &= 0  \quad &&\textrm{for } z=z_2,
\end{alignat}
\end{subequations}
where the length-scale $\gamma_j$ is given by
\begin{align}\label{gamma}
\begin{split}
	\gamma_j &= (-1)^{j+1} \frac{\unit z\cdot\left(\vec k \times \grad g_j\right)}{\unit z\cdot\left(\vec k \times \grad f\right)}\\ &= (-1)^{j+1}  \left( \frac{\alpha_j \, k}{\beta \, k_x} \right) \, \sin\left(\Delta \theta_j\right)
	\end{split}
\end{align}
 where $\alpha_j = |\grad{g_j}|$ and $\Delta \theta_j$ is the angle between the wavevector $\vec k$ and $\grad g_j$ measured counterclockwise from $\vec k$. The parameter $\gamma_j$ depends only on the direction of the wavevector $\vec k$ and not its magnitude $k = |\vec k|$. If $\gamma_j = 0$, then the $j$th boundary condition can be written as a $\lambda$-independent boundary condition [as in the upper boundary condition at $z=z_{2}$ of the eigenvalue problem \eqref{Rhines-eigen_bottom}]. For now, we assume that $\gamma_1\neq 0$. 
 
 Since the eigenvalue, $\lambda$, appears in the differential equation and  one boundary condition in the eigenvalue problem \eqref{Rhines-eigen_bottom}, the eigenvalue problem takes place in $L^2\oplus \C$.
 
 \subsubsection{Characterizing the eigen-solutions}
The following is obtained by applying the theory summarized in appendix A to the eigenvalue problem \eqref{Rhines-eigen_bottom}.\footnote{To apply the theory of \cite{yassin_normal_nodate}, summarized in Appendix A, let $\tilde \lambda = \lambda - k^2$ be the eigenvalue in place of $\lambda$; the resulting eigenvalue problem for $\tilde \lambda$ will then satisfy the positiveness conditions, equations \eqref{left-definite-1} and \eqref{left-definite-2}, of Appendix A.}
 
The eigenvalue problem \eqref{Rhines-eigen_bottom} has a countable infinity of eigenfunctions $G_0,\,G_1,\, G_2, \dots$ with ordered and distinct non-zero eigenvalues $\lambda_n$ satisfying
\begin{equation}\label{eigen-order}
	\lambda_0 < \lambda_1 < \lambda_2 < \cdots \rightarrow \infty.
\end{equation}
The inner product $\inner{\cdot}{\cdot}$ induced by the eigenvalue problem 
\eqref{Rhines-eigen_bottom}
is 
\begin{equation}\label{Rhines-inner}
	\inner{F}{G}= \frac{1}{H}\left( \intz F \, G \, \mathrm{d}z + \gamma_1 \, F(z_1)\, G(z_1) \right),
\end{equation}
which depends on the direction of the horizontal wavevector $\vec k$ through $\gamma_1$. Moreover, $\gamma_1$ is not necessarily positive\footnote{That $\gamma_1$ is not positive prevents us from applying the eigenvalue theory outlined in the appendix of \cite{smith_surface-aware_2012}.}, with one consequence being that some functions $G$ may have a negative square, $\inner{G}{G}<0$. Orthonormality  of the modes $G_n$ then takes the form
\begin{equation}
	\pm \delta_{mn} = \inner{G_m}{G_n},
\end{equation}
where at most one mode, $G_n$, satisfies $\inner{G_n}{G_n}=-1$. The eigenfunctions $\{G_n\}_{n=0}^\infty$ form an orthonormal basis of $L^2\oplus \C$ under the inner product $\eqref{Rhines-inner}$.

Appendix A provides the following inequality
\begin{equation}
	\left(k^2 + \lambda_n \right) \inner{G_n}{G_n} > 0
\end{equation}
which, using the dispersion relation \eqref{eigenvalue}, implies that modes $G_n$ with $\inner{G_n}{G_n} >0$ correspond to waves with a westward phase speed while modes $G_n$ with $\inner{G_n}{G_n}<0$ correspond to waves with an eastward phase speed (assuming $\beta>0$).

\begin{figure*}
	\centerline{\includegraphics[width=27pc]{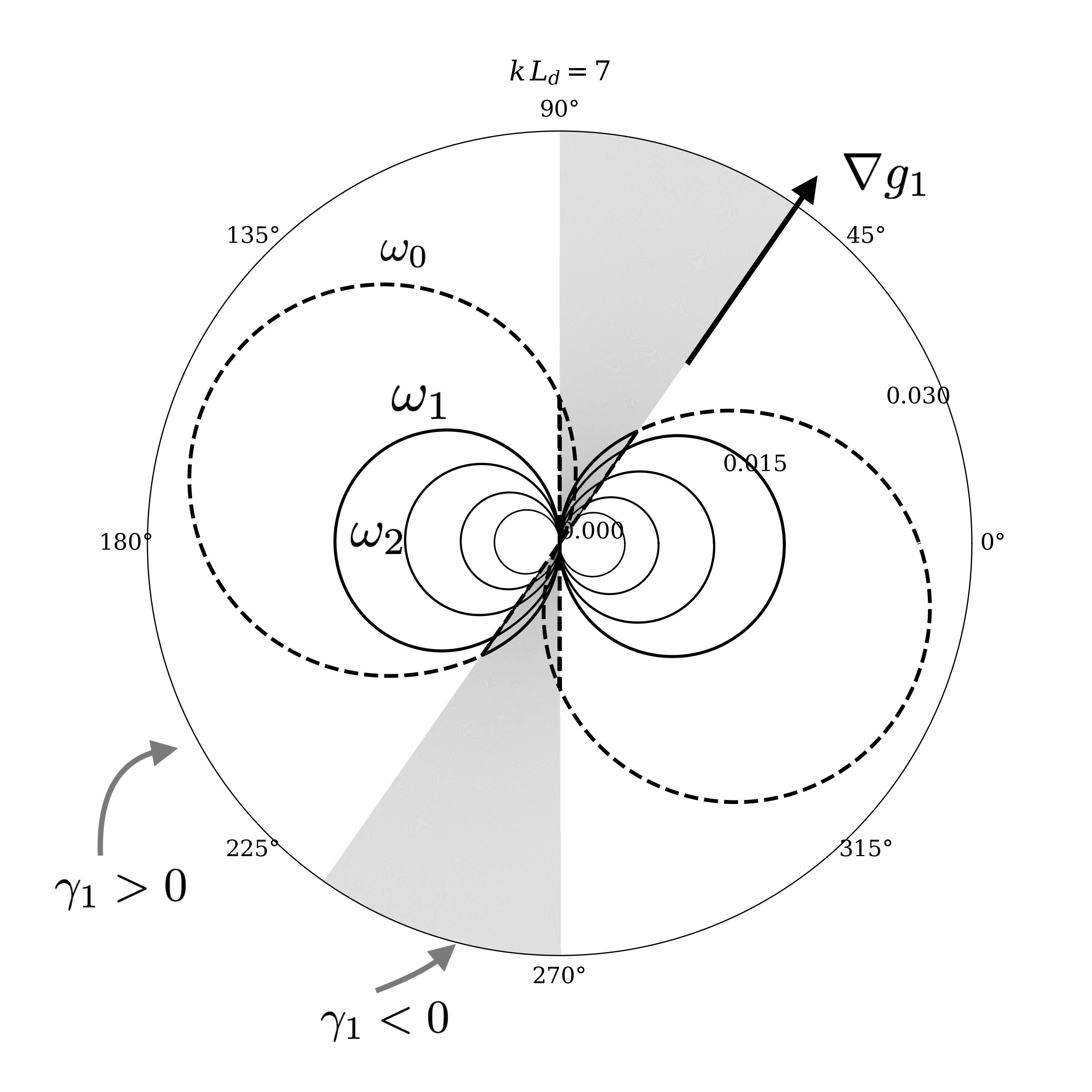}}
	\caption{Polar plots of the absolute value of the non-dimensional angular frequency $|\omega_n|/(\beta L_d)$ of the first five modes from section \ref{S-linear}\ref{SS-rhines-bottom} as a function of the wave propagation direction $\vec k/|\vec k|$ for a horizontal wavenumber given by $k\, L_d = 7$ in constant stratification. The dashed line corresponds to $\omega_0$, this mode becomes boundary trapped at large wavenumbers $k = |\vec k|$. The remaining modes, $\omega_n$ for $n=1,2,3,4$, are shown with solid lines. White regions are angles where $\gamma_1>0$. All Rossby waves with a propagation direction lying in the white region have negative angular frequencies $\omega_n$ and so have a westward phase speed. Gray regions are angles where $\gamma_1<0$. Here, $\omega_0$ is positive while the remaining angular frequencies $\omega_n$ for $n>0$ are negative. Consequently, in the gray regions, $\omega_0$ corresponds to a Rossby wave with an eastward phase speed whereas the remaining Rossby waves have westward phase speeds.  The lower boundary buoyancy gradient, proportional to $\grad g_1$, points towards $55^\circ$ and corresoponds to a bottom slope of $|\grad h_1| = 1.5\times 10^{-5}$ leading to $\gamma_1/H = 0.15$. The remaining parameters are as in figure \ref{F-baroclinic_angle}.}
	\label{F-angle1}
\end{figure*}

We distinguish the following cases depending on the sign of $\gamma_1$. In the following, we assume $k\neq0$.
\begin{itemize}
	\item [i.] $\gamma_1 > 0$. All eigenvalues satisfy $\lambda_n > -k^2$, all modes satisfy $\inner{G_n}{G_n}>0$, and all waves propagate westward. The $n$th mode, $G_n$, has $n$ internal zeros \citep{binding_sturmliouville_1994}. See the regions in white in figure \ref{F-angle1}.
	
	\item [ii.]  $\gamma_1 < 0$. There is one mode, $G_0$, with a negative square, $\inner{G_0}{G_0} <0$, corresponding to an eastward propagating wave. The eastward propagating wave nevertheless travels pseudowestward (to the left of the upslope direction for $f_0>0$). The associated eigenvalue, $\lambda_0$, satisfies $\lambda_0 < -k^2$. The remaining modes, $G_n$ for $n>1$, have positive squares, $\inner{G_n}{G_n} >0$, corresponding to westward propagating waves and have eigenvalues, $\lambda_n$, satisfying $\lambda_n>-k^2$. Both $G_0$ and $G_1$ have no internal zeros whereas the remaining modes, $G_n$, have $n-1$ internal zeros for $n>1$ \citep{binding_sturmliouville_1994}. See the stippled regions in figures \ref{F-angle1}.
\end{itemize}


To elucidate the meaning of $\lambda_n < -k^2$, note that a pure surface quasigeostrophic mode\footnote{A pure surface quasigeostrophic mode is the mode found after setting $\beta =0$ with an upper boundary at $z_2=\infty$.}  has $\lambda = -k^2$. Thus $\lambda_0 < -k^2$ means that the bottom-trapped mode decays away from the boundary more rapidly than a pure surface quasigeostrophic wave. Indeed, the limit of $\lambda_0 \rightarrow -\infty$ yields the bottom step-mode \eqref{step-mode} of the previous subsection.

\begin{figure}
	\noindent \includegraphics[width=\columnwidth]{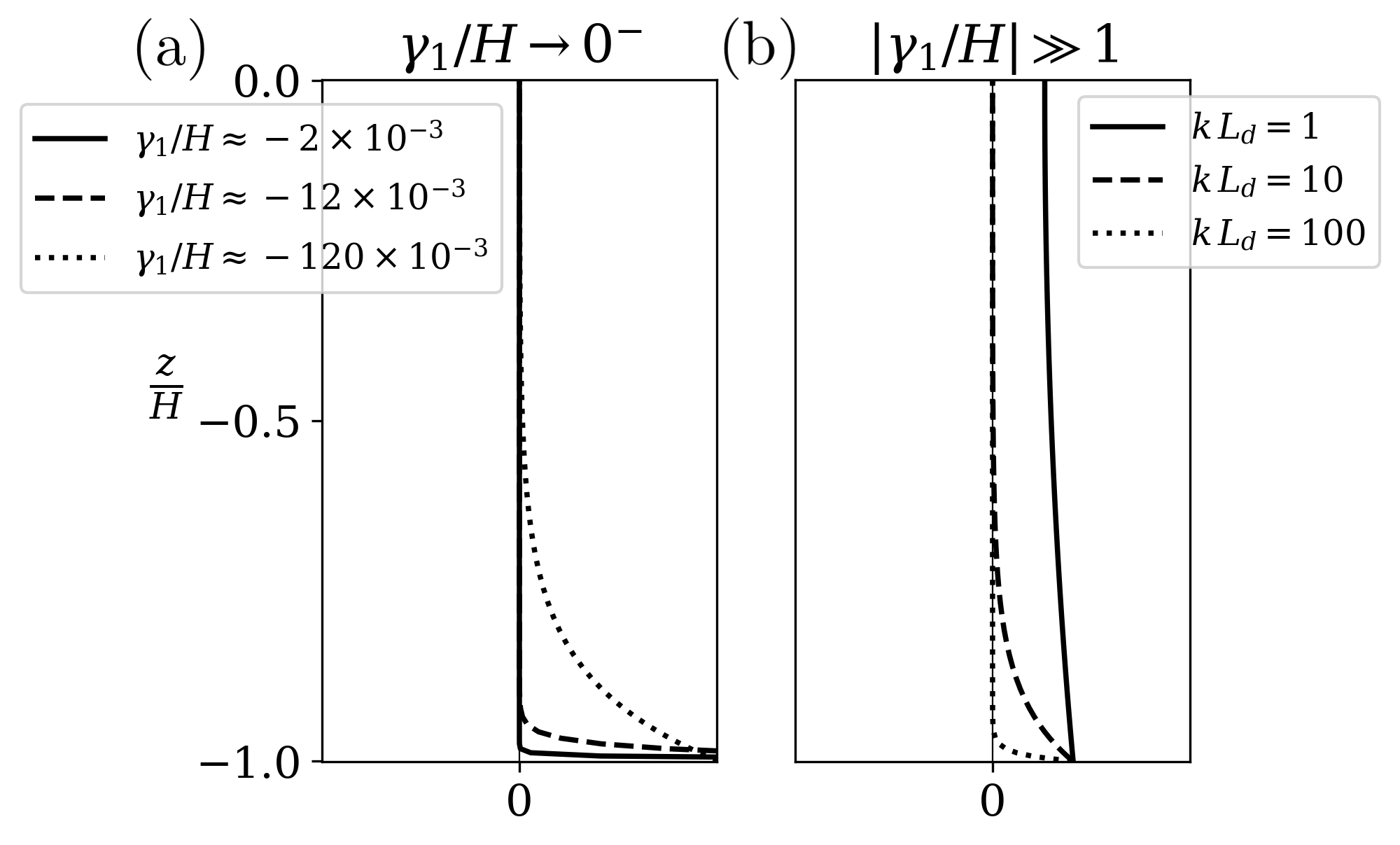}
	\caption{The two limits of the boundary-trapped surface quasigeostrophic waves, as discussed in section \ref{S-linear}\ref{SS-rhines}. (a) Convergence to the step mode given in equation \eqref{step-mode} with $j=1$ as $\gamma_1 \rightarrow 0^-$ for three values of $\gamma_1$ at a  wavenumber $k=|\vec k|$ given by $k \, L_d =1$. The phase speed approaches zero in the limit $\gamma_1 \rightarrow 0^{-}$. (b) Here, $\gamma_1/H \approx 10$ for the three vertical structures $G_n$ shown. Consequently, the bottom trapped wave has $\lambda \approx -k^2$ and the phase speeds are large. The vertical structure, $G$, for three values of $k \, L_d$ are shown, illustrating the dependence on $k$ of this mode, which behaves as a boundary-trapped exponential mode with an $\mathrm{e}$-folding scale of $|\lambda|^{-1/2} = k^{-1}$. In both (a) and (b), the wave propagation direction $\theta = 260^\circ$. All other parameters are identical to figure \ref{F-angle1}.}
	\label{F-step}
\end{figure}

The step-mode limit is obtained as $\gamma_1 \rightarrow 0^-$. This limit is found as either $|\grad g_1| \rightarrow 0$ for propagation directions in which $\gamma_1 < 0$ or as $\vec k$ becomes parallel or anti-parallel to $\grad g_1$ (whichever limit satisfies $\gamma_1 \rightarrow 0^-$). In this limit, we obtain a step-mode exactly confined at the boundary (that is, $|\lambda|^{-1/2}=0$) with zero phase speed [see figure \ref{F-step}(a)]. The remaining modes then satisfy the isentropic boundary condition
\begin{equation}\label{non-zero-boundary}
	\left(\frac{f_0^2}{N^2} \, \d{G_n}{z}\right)\Bigg |_{z=z_1}= 0.
\end{equation}

The other limit is that of $|\gamma_1| \rightarrow \infty$ which is obtained as the buoyancy gradient becomes large, $|\grad g_1| \rightarrow \infty$. In this limit, the eigenvalue $\lambda_0 \rightarrow -k^2$ [see figure \ref{F-step}(b)]. Moreover, the phase speed of the bottom-trapped wave becomes infinite, an indication that the quasigeostrophic approximation breaks down. Indeed, the large buoyancy gradient limit corresponds to steep topographic slopes and so we obtain the topographically-trapped internal gravity wave of \cite{rhines_edge_1970}, which has an infinite phase speed in quasigeostrophic theory. The remaining modes then satisfy the vanishing pressure boundary condition
\begin{align}\label{zero-boundary}
	G(z_1) = 0
\end{align}
as in the surface modes of \cite{de_la_lama_vertical_2016} and \cite{lacasce_prevalence_2017}.



\subsubsection{The general time-dependent solution}

At some wavevector $\vec k$, the observed vertical structure now has the form
\begin{equation}\label{rhines-vertical}
	\Psi(z)  = \psi_{\vec k} (z,t=0) 
\end{equation}
where $\Psi$ is a twice continuously differentiable function satisfying $d\Psi(z_2)/dz=0$. For such functions we can write (see appendix A)
\begin{equation}
	\Psi = \sum_{n=0}^\infty \frac{\inner{\Psi}{G_n}}{\inner{G_n}{G_n}} G_n.
\end{equation}
so that the time-evolution is
\begin{equation}\label{rhines-evolution}
	\psi_{\vec k} (z,t) = \sum_{n=0}^\infty \frac{\inner{\Psi}{G_n}}{\inner{G_n}{G_n}} G_n(z) \, \mathrm{e}^{-\mathrm{i}\omega_n t}.
\end{equation}
Again, it is the above expression, which is valid only in linear theory with a quiescent background state, that gives the generalized Rhines modes $G_n$ physical meaning. Outside the linear theory of this section, the generalized Rhines modes do not have any physical interpretation and instead merely serve as a mathematical basis for $L^2\oplus \C$.

Recall from section \ref{S-linear}\ref{SS-trad} that an expansion of a step-mode \eqref{step-mode} in terms of the baroclinic modes $\{F_n\}_{n=0}^\infty$  produces a series that is identically zero. It follows that the step-modes are independent of the baroclinic modes\textemdash they constitute independent degrees of freedom. However, with the inclusion of bottom boundary dynamics, we may now expand the bottom step-mode, $F^\mathrm{step}_1(z)$, in terms of the $L^2\oplus\C^1$ modes, $\{G_n\}_{n=0}^\infty$, with the expansion given by
\begin{equation}
	F^\mathrm{step}_1(z) = \frac{\gamma_1}{H} \sum_{n=0}^\infty \frac{G_n(z_1)}{\inner{G_n}{G_n}} G_n(z).
\end{equation}

\begin{figure*}
	\centerline{\includegraphics[width=27pc]{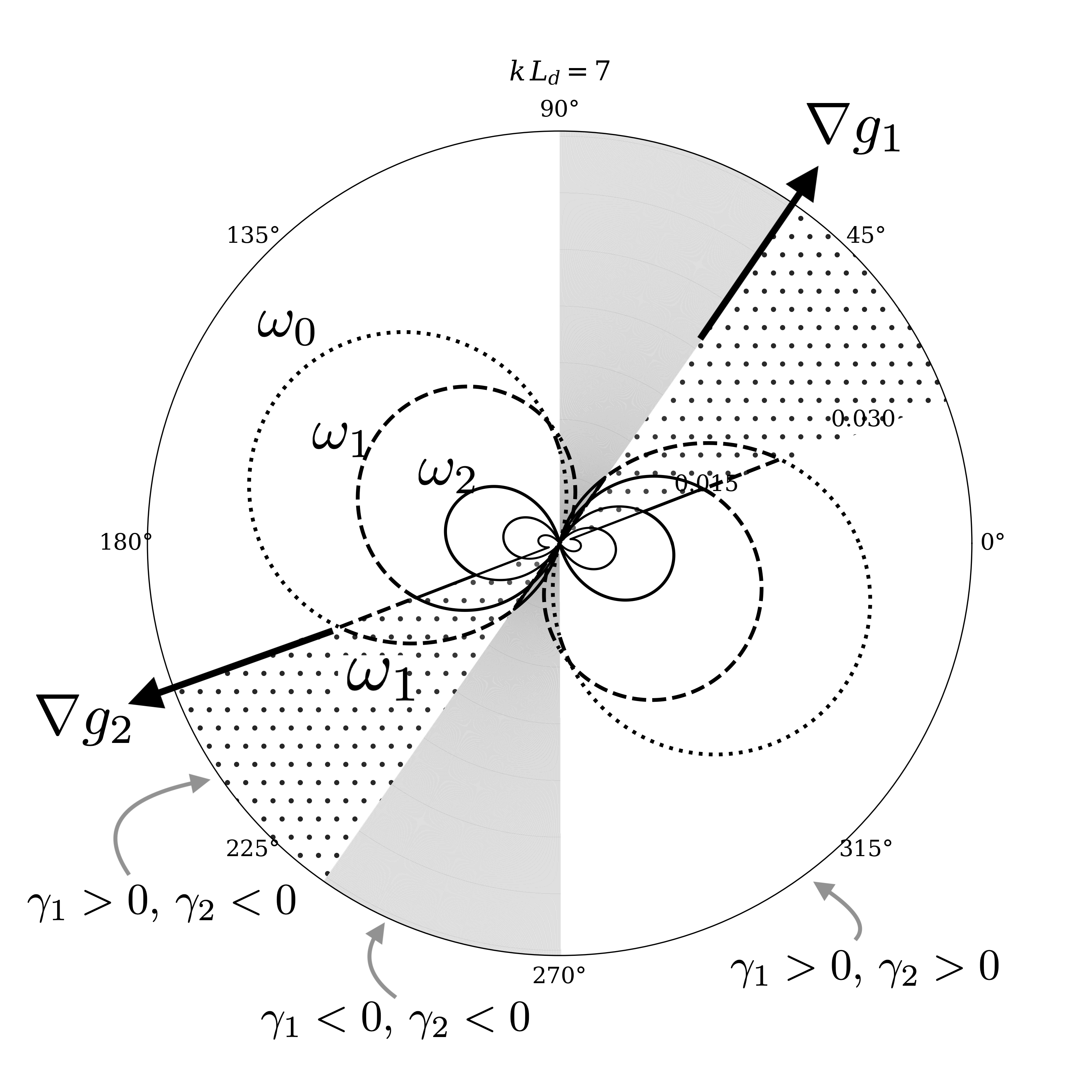}}
	\caption{As in figure \ref{F-angle1} but now with an upper slope $|\grad h_2| = 10^{-5}$ in the direction $200^\circ$ in addition to the bottom slope in figure \ref{F-angle1}. The upper slope corresponds to $\gamma_2/H =0.1$. The dotted line corresponds to $\omega_0$, the dashed line to $\omega_1$, with these two modes becoming boundary trapped at large wavenumbers $k$. The remaining modes, $\omega_n$ for $n=2,3,4$, are shown with solid lines. White regions are angles where $\gamma_1>0$ and $\gamma_2>0$. All Rossby waves with a propagation direction lying in the white region have negative angular frequencies $\omega_n$ and so have a westward phase speed. Gray regions are angles where $\gamma_1<0$ and $\gamma_2<0$. The two gravest angular frequencies $\omega_0$ and $\omega_1$ are both positive while the remaining angular frequencies $\omega_n$ for $n>1$ are negative. Consequently, in the gray regions, $\omega_0$ and $\omega_1$ each correspond to a Rossby waves with an eastward phase speed whereas the remaining Rossby waves have westward phase speeds . Stippled regions are angles where $\gamma_1>0$ and $\gamma_2<0$. In the stippled region, $\omega_0$ is positive and has an eastward phase speed. The remaining Rossby waves in the stippled region have negative angular frequencies and have westward phase speeds.}
	\label{F-angle2}
\end{figure*}

\subsection{The generalized Rhines problem}\label{SS-rhines}

The general problem with topography at both the upper and lower boundaries is
\begin{subequations}\label{Rhines-eigen}
\begin{align}	
		\label{Rhines-eigen-interior}		
		- \d{}{z} \left(\frac{f_0^2}{N^2} \d{G}{z}\right) = \lambda \, G \quad &\textrm{for } z\in(z_1,z_2)\\
		\label{Rhines-eigen-boundary}
			-k^2 G + (-1)^j \gamma_j^{-1} \left( \frac{f_0^2}{N^2}\d{G}{z} \right)= \lambda \, G \quad &\textrm{for } z=z_j,
\end{align}
\end{subequations}
for $j=1,2$, where the length-scale $\gamma_j$ is given by equation \eqref{gamma}. As the eigenvalue, $\lambda$, appears in both boundary conditions, the eigenvalue problem \eqref{Rhines-eigen} takes place in $L^2\oplus \C^2$. The inner product now has the form
\begin{equation}\label{Rhines-inner-general}
	\inner{F}{G}= \frac{1}{H}\left( \intz F \, G \, \mathrm{d}z + \sum_{j=1}^2 \gamma_j \, F(z_j)\, G(z_j) \right)
\end{equation}
which reduces to equation \eqref{Rhines-inner} when $\gamma_2=0$. Under this inner product, the eigenfunctions $\{G_n\}_{n=0}^\infty$ form a basis of $L^2\oplus \C^2$.

\begin{figure*}
	\noindent \includegraphics[width=\textwidth]{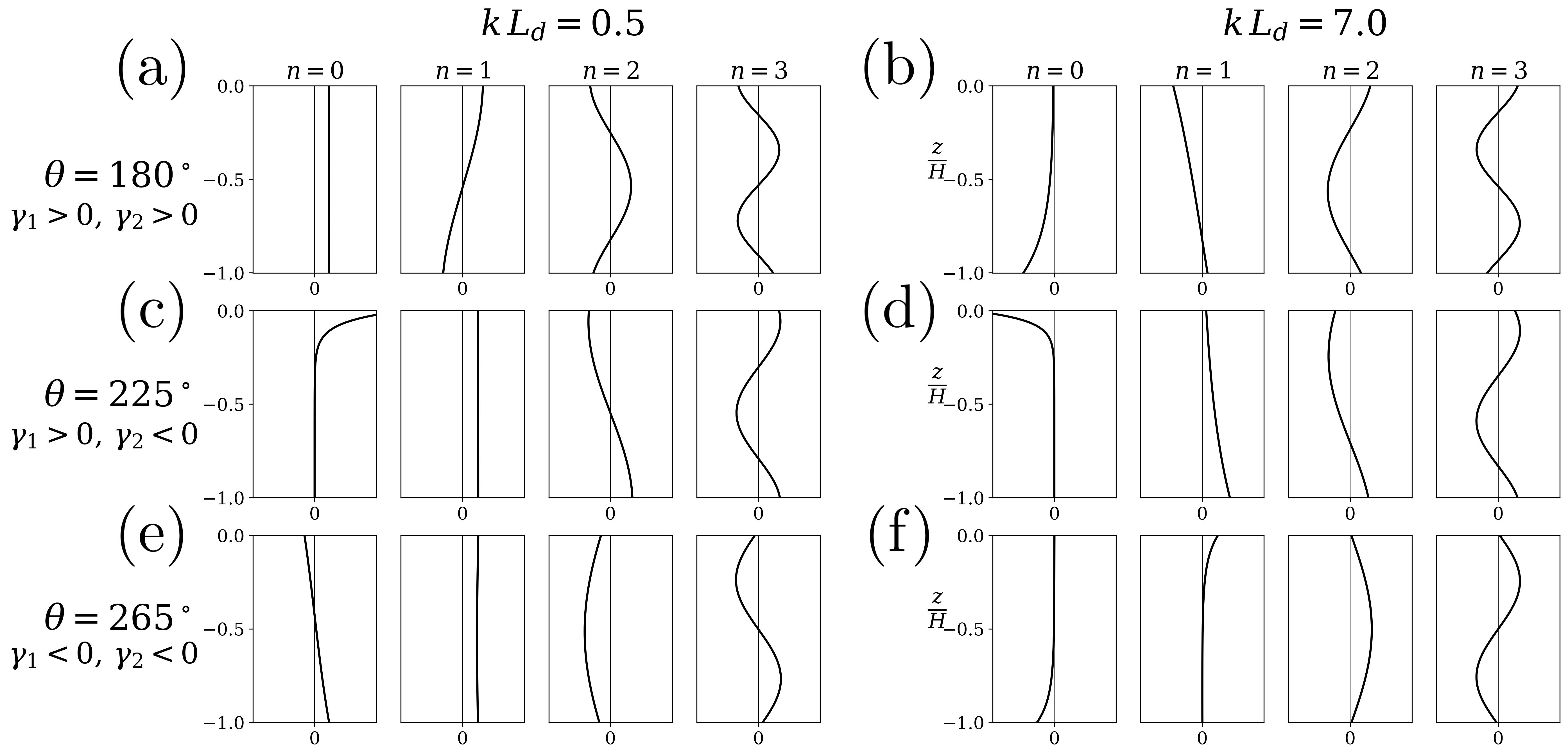} \\
	\caption{This figure illustrates the dependence of the vertical structure $G_n$ of the streamfunction to the horizontal wavevector $\vec k$ as discussed in section \ref{S-linear}\ref{SS-rhines} . Three propagation directions are shown $\theta = 180^\circ, \, 225^\circ, \, 265^\circ$ and correspond to the rows in the figure [e.g., the row containing (a) and (b) are the vertical structures of waves at $\theta = 180^\circ$]; two wavenumbers $k\, L_d = 0.5, 7$ are shown (where $k=|\vec k|$) and they correspond to the columns in the above figure [e.g., (b), (d) and (f) are the vertical structure of waves with $k \, L_d = 7$]. The parameters for the above figure are identical to figure \ref{F-angle1}. We emphasize two features in this figure. First, note how the boundary modes ($n=0,1$) are typically only boundary-trapped at small horizontal scales (i.e., for $k \, L_{d} = 7$). At larger horizontal scales, we typically obtain a depth-independent mode along with another mode with large-scale features in the vertical. Second, note that for $\gamma_1,\gamma_2>0$, as in panels (a) and (b), the $n$th mode has $n$ internal zeros, as in Sturm-Liouville theory; for $\gamma_1>0,\gamma_2<0$, as in panels (c) and (d), the first two modes $(n=0,1)$ have no internal zeros; and for $\gamma_1,\gamma_2<0$, the zeroth mode $G_0$ has one internal zero, the first and second modes, $G_1$ and $G_2$ have no internal zeros, and the third mode $G_2$ has one internal zero. The zero-crossing for the $n=0$ mode in panel (f) is difficult to observe because the amplitude of $G_0$ is small near the zero-crossing.
	}
	\label{F-little}
\end{figure*}
 
There are now three cases depending on the signs of $\gamma_1$ and $\gamma_2$ and as depicted in figures \ref{F-angle2} and \ref{F-little}. In the following, we assume $k\neq0$. 
\begin{itemize}
	\item [i.] $\gamma_1 > 0$ and $\gamma_2 > 0$. Corresponds to case (i) in section \ref{S-linear}\ref{SS-rhines-bottom}. See the regions in white in figure \ref{F-angle2} and plots (a) and (b) in figure \ref{F-little}.
	
	\item [ii.] $\gamma_1 \, \gamma_2 < 0$. This corresponds to case (ii) in section \ref{S-linear}\ref{SS-rhines-bottom}. See the stippled regions in figure  and \ref{F-angle2} and plots (c) and (d) in figure \ref{F-little}.
	
	\item [iii.] $\gamma_1 < 0$ and $\gamma_2 < 0$. There are two modes $G_0$ and $G_1$ with negative squares, $\inner{G_n}{G_n}<0$, that propagate eastward and have eigenvalues, $G_n$, satisfying $G_n<-k^2$ for $n=1,2$. The remaining modes, $G_n$, for $n>1$ have positive squares, $\inner{G_n}{G_n}>0$, propagate westward, and have eigenvalues, $\lambda_n$, satisfying $\lambda_n>-k^2$. The zeroth mode, $G_0$, has one internal zero, the first and second modes, $G_1$ and $G_2$, have no internal zeros, and the remaining modes, $G_n$, have $n-2$ internal zeros for $n>2$ \citep{binding_left_1999}. See the shaded regions in figures \ref{F-angle1} and \ref{F-angle2} and panels (e) and (f) in figure \ref{F-little}.
\end{itemize}

\subsection{The vertical velocity eigenvalue problem}\label{SS-vertical-velocity}

Let $\hat w(z) = \hat w_0 \, \chi(z)$ where $\chi(z)$ is a non-dimensional function. For the Rossby waves with isentropic boundaries of section \ref{S-linear}\ref{SS-trad} (the traditional baroclinic modes), the corresponding vertical velocity modes satisfy
\begin{equation}\label{vertical-baroclinic-interior}
	-\dd{\chi}{z} = \lambda  \left(\frac{N^2}{f_0^2}\right) \chi
\end{equation}
with vanishing vertical velocity boundary conditions
\begin{equation}\label{vertical-baroclinic-boundary}
	\chi(z_j) = 0
\end{equation}
(see appendix B for details). The resulting modes $\{\chi_n\}_{n=0}^\infty$ form an orthonormal basis of $L^2$ with orthonormality given by
\begin{equation}
	\delta_{mn} = \frac{1}{H} \intz \chi_m\, \chi_n \left(\frac{N^2}{f_0^2}\right)  \mathrm{d}z.
\end{equation}
One can obtain the eigenfunctions, $\chi_n$, by solving the eigenvalue problem \eqref{vertical-baroclinic-interior}\textendash\eqref{vertical-baroclinic-boundary} or by differentiating the streamfunction modes $F_n$ according to equation \eqref{psiz-w}. 

\subsubsection*{Quasigeostrophic boundary dynamics}

As seen earlier, boundary buoyancy gradients activate  boundary dynamics in the quasigeostrophic problem. In this case, boundary conditions for the quasigeostrophic vertical velocity problem \eqref{vertical-baroclinic-interior} become
\begin{equation}\label{vertical-velocity-boundary}
	- (-1)^j \, \gamma_j \, k^2 \, \d{\chi}{z}\Big|_{z_j} = \lambda \left[\chi|_{z_j} + (-1)^j \, \gamma_j \, \d{\chi}{z}\Big|_{z_j} \right]
\end{equation}
(see the appendix B). The resulting modes $\{\chi_n\}_{n=0}^\infty$ satisfy a peculiar orthogonality relation given by equation \eqref{vertical-velocity-ortho}.

\begin{figure*}
	\noindent \includegraphics[width=\textwidth]{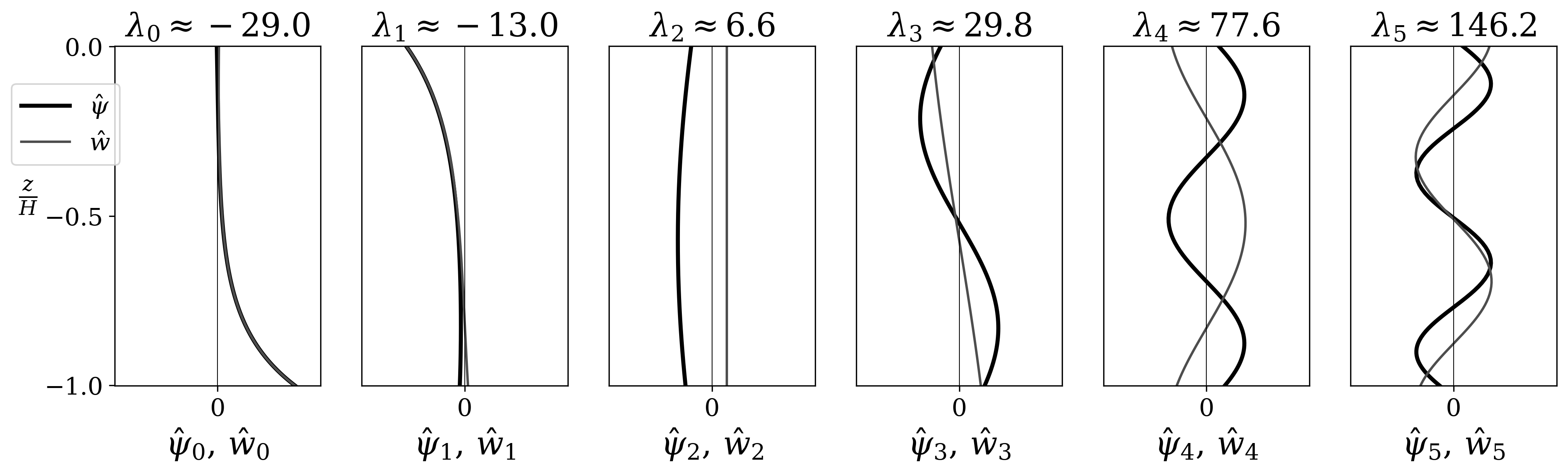} \\
	\caption{The first six vertical velocity normal modes $\chi_n$ (thin grey lines) and streamfunction normal modes $G_n$ (black lines) (see section \ref{S-linear}\ref{SS-vertical-velocity}). The propagation direction is $\theta = 75^\circ$ with a wavenumber of $k\,L_d = 2$. The remaining parameters are as in figure \ref{F-angle1}. Note that $\chi_n$ and $G_n$ are nearly indistinguishable from the boundary-trapped modes $n=0,1$ while they are related by a vertical derivative for the internal modes $n>1$. The eigenvalue in the figure is non-dimensionalized by the deformation radius $L_d$.}
	\label{F-vertical-velocity}
\end{figure*}


\section{Eigenfunction expansions}\label{S-expansions}

Motivated by the Rossby waves of the previous section, we now investigate various sets of normal modes for quasigeostrophic theory. Let $\{F_n\}_{n=0}^\infty$ be a collection of $L^2$ normal modes, and assume $\psi_{\vec k}(z,t)$ is twice continuously differentiable in $z$. Define the eigenfunction expansion $\psi_{\vec k}^\textrm{exp}$ of $\psi$ by
\begin{equation}\label{F-expandion}
	\psi_{\vec k}^\textrm{exp}(z,t) = \sum_{n=0}^\infty \psi_{\vec k n}(t) \, F_n(z)
\end{equation}
where 
\begin{equation}
	\psi_{\vec k n} = \left[\psi_{\vec k},F_n\right].
\end{equation}
Because $\{F_n\}_{n=0}^\infty$ is a basis of $L^2$, the eigenfunction expansion $\psi_{\vec k}^\textrm{exp}$ satisfies \citep[e.g.,][] {brown_fourier_1993}
\begin{equation}\label{mean-square-equality}
	\intz |\psi_{\vec k}(z) - \psi_{\vec k}^\textrm{exp}(z)|^2 \mathrm{d}z = 0.
\end{equation}
Significantly, the vanishing of the integral \eqref{mean-square-equality} does not imply $\psi_{\vec k} = \psi_{\vec k}^\textrm{exp}$ because the two functions can still differ at some points $z \in [z_1,z_2]$.

In the following, we will only consider eigenfunctions expansions that diagonalize the energy and potential enstrophy integrals of section \ref{S-phase}\ref{subsection:decomposing-energy-enstrophy}.

\subsection{The four possible $L^2$ modes}\label{SS-L2-modes}

There are only four $L^2$ bases in quasigeostrophic theory that diagonalize the energy and potential enstrophy integrals. All four sets of corresponding normal modes satisfy the differential equation
\begin{equation}
	- \d{}{z} \left(\frac{f_0^2}{N^2} \d{F}{z} \right) = \lambda \, F 
	 \quad z \in (z_1,z_2),
\end{equation}
but differ in boundary conditions according to the following (recall that $z_{1}$ is the bottom and $z_{2}$ the surface). 
\begin{itemize}
	\item \textit{Baroclinic modes}: Vanishing vertical velocity at both boundaries (Neumann),
		\begin{equation}
			\d{F(z_1)}{z} =0, \quad \d{F(z_2)}{z} = 0.
		\end{equation}
		\item \textit{Anti-baroclinic modes}: Vanishing pressure\footnote{Recall that the geostrophic streamfunction $\psi$ is proportional to pressure \cite[e.g.,][section 5.4]{Vallis2017}.} at both boundaries (Dirichlet),
		\begin{equation}
			F(z_1)=0, \quad F(z_2) = 0.
		\end{equation} 
	\item \textit{Surface modes}: (mixed Neumann/Dirichlet)
		\begin{equation}
			F(z_1) = 0, \quad \d{F(z_2)}{z} = 0.
		\end{equation}
	\item \textit{Anti-surface modes}: (mixed Neumann/Dirichlet)
		\begin{equation}
			\d{F(z_1)}{z} = 0, \quad F(z_2) = 0.
		\end{equation}
\end{itemize}
All four sets of modes are missing two modes. Each boundary condition of the form
\begin{equation}
	\d{F(z_j)}{z} = 0,
\end{equation}
implies a missing step-mode while a boundary condition of the form
\begin{equation}
	F(z_j) = 0,
\end{equation}
implies a missing boundary-trapped exponential mode [see the $\gamma_1\rightarrow \infty$ limit leading to equation \eqref{zero-boundary}].

\subsection{Expansions with $L^2$ modes}\label{SS-expansions-L2}

We here examine the pointwise convergence and the term-by-term differentiability of eigenfunction expansions in terms of $L^2$ modes. 
These properties of $L^2$ Sturm-Liouville expansions may be found in \cite{brown_fourier_1993} and \cite{levitan_introduction_1975}.\footnote{In particular, chapters 1 and 8 in \cite{levitan_introduction_1975} show that eigenfunction expansions have the same pointwise convergence and differentiability properties as the Fourier series with the analogous boundary conditions. The behaviour of Fourier series is discussed in \cite{brown_fourier_1993}.}

\subsubsection{Pointwise equality on $[z_1,z_2]$}

For all four sets of $L^2$ modes, if $\psi_{\vec k}$ is twice continuously differentiable in $z$, we obtain pointwise equality in the interior  
\begin{equation}
	\psi_{\vec k}(z) = \psi_{\vec k}^\textrm{exp}(z) \quad \textrm{for } z\in(z_1,z_z).
\end{equation}
The behaviour at the boundaries depends on the boundary conditions the modes $F_n$ satisfy. If the $F_n$ satisfy the vanishing pressure boundary condition at the $j$th boundary
\begin{equation}
	F_n(z_j) = 0 
\end{equation}
then
\begin{equation}
	\psi_{\vec k}^\textrm{exp}(z_j) = 0
\end{equation}
regardless of the values of $\psi_{\vec k}(z_j)$. It follows that $\psi_{\vec k}^\textrm{exp}$ will be continuous over $(z_1,z_2)$ and will generally have a jump discontinuity at the boundaries [unless $\psi_{\vec k}(z_j)=0$ for $j=1,2$]. In contrast, if the $F_n$ satisfy a zero vertical velocity boundary condition at the $j$th boundary
\begin{equation}
	\d{F_n(z_j)}{z} = 0
\end{equation} 
then 
\begin{equation}
	 \psi_{\vec k}(z_j) = \psi_{\vec k}^\textrm{exp}(z_j).
\end{equation}
Consequently, of the four sets of $L^2$ modes, only with the baroclinic modes do we obtain the pointwise equality $\psi_{\vec k}(z) = \psi_{\vec k}^\textrm{exp}(z)$ on the \emph{closed} interval $[z_1,z_2]$. 

However, even though $\psi_{\vec k}^\textrm{exp}$ converges pointwise to $\psi_{\vec k}$ when the baroclinic modes are used, we are unable to represent the corresponding velocity $w_{\vec k}$ in terms of the vertical velocity baroclinic modes since the modes vanish at both boundaries. Analogous considerations show that only the anti-baroclinic vertical velocity modes (see appendix B) can represent arbitrary vertical velocities.

\subsubsection{Differentiability of the series expansion}

Although  we obtain pointwise equality on the whole interval $[z_1,z_2]$ with the streamfunction baroclinic modes, we have lost two degrees of freedom in the expansion process. Recall that the degrees of freedom in the quasigeostrophic phase space are determined by the potential vorticity. The volume potential vorticity, $q_{\vec k}$, is associated with the $L^2$ degrees of freedom while the surface potential vorticities, $r_{1\vec k}$ and $r_{2 \vec k}$, are associated with the $\C^2$ degrees of freedom.

The series expansion $\psi_{\vec k}^\textrm{exp}$ of $\psi_{\vec k}$ in terms of the baroclinic modes is differentiable in the interior $(z_1,z_2)$. Consequently, we can differentiate the $\psi_{\vec k}^\textrm{exp}$ series for $z\in(z_1,z_2)$ to recover $q_{\vec k}$, that is,
\begin{equation}\label{q-trad-expansion}
	q_{\vec k}  = \sum_{n=0}^\infty  q_{\vec k n}\, F_n
\end{equation}
where 
\begin{equation}\label{qn-trad-expansion}
	q_{\vec k n} = - (k^2 + \lambda_n) \, \psi_{\vec k n}.
\end{equation}
However, $\psi_{\vec k}^\textrm{exp}$ is not differentiable at the boundaries, $z=z_1,z_2$, so we are unable to recover the surface potential vorticities, $r_{1 \vec k}$ and $r_{2 \vec k}$. Two degrees of freedom are lost by projecting onto the baroclinic modes.\footnote{To see that $\psi_{\vec k}^\textrm{exp}$ is non-differentiable at $z=z_1,z_2$, suppose that the series $\psi_{\vec k}^\textrm{exp}$ is differentiable and that $\mathrm{d}\psi_{\vec k}(z_j)/\mathrm{d}z \neq 0$ for $j=1,2$. But then $$ 0 \neq \d{\psi_{\vec k}(z_j)}{z} = \sum_{n=0}^\infty \psi_{\vec k n} \d{F_n(z_j)}{z} = 0,$$ which is a contradiction.}

The energy at wavevector $\vec k$ is indeed partitioned between the modes
\begin{align}\label{trad-energy}
	E_{\vec k} = \sum_{n=0}^\infty (k^2 + \lambda_n) \, \psi_{\vec k n}
\end{align}
and similarly for the potential enstrophy
\begin{align}\label{trad-enstrophy-expansion}
		Z_{\vec k} = \sum_{n=0}^\infty (k^2 + \lambda_n)^2 \psi_{\vec k n}.
\end{align}
However, as we have lost $r_{1 \vec k}$ and $r_{2 \vec k}$ in the projection process, the surface potential enstrophies $Y_{1 \vec k}$ and $Y_{2 \vec k}$, defined in equation \eqref{surface-enstrophy-rjk}, are not partitioned.

\subsection{Quasigeostrophic $L^2 \oplus \C^2$ modes}

Consider the eigenvalue problem
\begin{subequations}\label{SV-eigen}
\begin{align}
			- \d{}{z} \left(\frac{f_0^2}{N^2} \d{G}{z}\right) = \lambda \, G \quad \textrm{for } z\in(z_1,z_2)&\\
			-k^2 G + (-1)^j D_j^{-1} \, \left(\frac{f_0^2}{N^2}\d{G}{z}\right) = \lambda \, G \quad \textrm{for } z=z_j&
\end{align}
\end{subequations}
where $D_1$ and $D_2$ are non-zero real constants. This eigenvalue problem differs from the generalized Rhines eigenvalue problem \eqref{Rhines-eigen} in that $D_j$ are generally not equal to the $\gamma_j$ defined in equation \eqref{gamma}. The inner product $\inner{\cdot}{\cdot}$ induced by the eigenvalue problem \eqref{SV-eigen} is given by equation \eqref{Rhines-inner-general} with the $\gamma_j$ replaced by the $D_j$.

\cite{smith_surface-aware_2012} investigate an equivalent eigenvalue problem to \eqref{SV-eigen} and conclude that, when $D_1$ and $D_2$ are positive, the resulting eigenfunctions form a basis of $L^2\oplus \C^2 $. However, such a completeness result is insufficient for the Rossby wave problem of  section \ref{S-linear}\ref{SS-rhines}, in which case $D_j=\gamma_j$ and $\gamma_j$ can be negative. 


\subsection{Expansion with $L^2\oplus \C^2$ modes}

When $D_1,D_2$ in the eigenvalue problem \eqref{SV-eigen} are finite and non-zero, the resulting eigenmodes $\{G_n\}_{n=0}^\infty$ form a basis for the vertical structure phase space $L^2\oplus \C^2$. Thus, the projection 
\begin{equation}
	\psi_{\vec k}^\textrm{exp}(z) = \sum_{n=0}^\infty \psi_{\vec k n} \, G_n(z)
\end{equation}
where
\begin{equation}
	\psi_{\vec k n} =  \frac{\inner{\psi_{\vec k}}{G_n}}{\inner{G_n}{G_n}}
\end{equation}
is an \emph{equivalent} representation of $\psi_{\vec k}$. Not only do we have pointwise equality
\begin{equation}
	\psi_{\vec k}(z) = \psi_{\vec k}^\textrm{exp}(z) \quad \textrm{for } z\in[z_1,z_2],
\end{equation}
but the series $\psi_{\vec k}^\textrm{exp}$ is also differentiable on the \emph{closed} interval $[z_1,z_2]$ [the case of $D_j>0$ is due to \cite{fulton_two-point_1977} whereas the case of $D_j<0$ is due to \cite{yassin_normal_nodate}.]. Thus given $\psi_{\vec k}^\textrm{exp}$, we can differentiate to obtain both $q_{\vec k}$ and $r_{j \vec k}$ and thereby recover all quasigeostrophic degrees of freedom. Indeed, we have 
\begin{align}
	q_{\vec k}(z,t) &= \sum_{n=0}^\infty q_{\vec k n}(t) \, G_n(z)\\
	r_{j \vec k}(t) &= \sum_{n=0}^\infty r_{j \vec k n}(t) \, G_n(z_j)
\end{align}
where
\begin{align}
	q_{\vec k n} &= - (k^2+\lambda_n) \frac{\inner{\Psi}{G_n}}{\inner{G_n}{G_n}} \\
	r_{j \vec k n} &= D_j \, q_{\vec k n} 
\end{align}
for $j=1,2$. 

In addition, the energy, $E_{\vec k}$, volume potential enstrophy, $Z_{\vec k}$, and surface potential enstrophies, $Y_{1\vec k}$ and $Y_{2\vec k}$, are partitioned (diagonalized) between the modes
\begin{align}\label{Rhines-energy}
	E_{\vec k} &=  \sum_{n=0}^\infty (k^2 + \lambda_n) \psi_{\vec k n}, \\
	Z_{\vec k} + \frac{1}{H} \sum_{j=1}^2 \frac{1}{D_j} Y_{j\vec k} &= \sum_{n=0}^\infty (k^2 + \lambda_n)^2 \psi_{\vec k n}.
\end{align}

\section{Discussion}\label{S-discussion}

The traditional baroclinic modes are useful since they are the vertical structures of linear Rossby waves in a resting ocean and they can be used for wave-turbulence studies such as in \cite[e.g.,][]{hua_numerical_1986,smith_scales_2001}. Therefore, any basis we choose should not only be complete in $L^2\oplus \C^2$, but should also represent the vertical structure of Rossby waves in the linear (quiescent ocean) limit. Such a basis would then amenable to wave-turbulence arguments and can permit a dynamical interpretation of field observations. The basis suggested by \cite{smith_surface-aware_2012} does not correspond to Rossby waves in the linear limit. It is a mathematical basis with two-independent parameters $D_1,D_2>0$ that diagonalizes the energy and potential enstrophy integrals. 

The Rhines modes of section \ref{S-linear}\ref{SS-rhines-bottom} offer a basis of $L^2\oplus \C$ that corresponds to Rossby wave over topography in the linear limit. These Rhines modes do not contain any free parameters. Indeed, if we set $D_2=0$ in the eigenvalue problem \eqref{SV-eigen} and let $D_1=\gamma_1$, we then obtain the Rhines modes. Note that since  $D_1=\gamma_1=\gamma_1(\vec k)$ may be negative, the \cite{smith_surface-aware_2012} modes do not apply. Instead, the case of negative $D_j$ is examined in this article and in \cite{yassin_normal_nodate}.

However, the Rhines modes, as a basis of $L^2\oplus \C$ are not a basis of the whole vertical structure phase space $L^2\oplus \C^2$ since they exclude surface buoyancy anomalies at the upper boundary. To solve this problem, we can use the modes of the eigenvalue problem \eqref{SV-eigen} with $D_1=\gamma_1$ but leaving $D_2$ arbitrary as in \cite{smith_surface-aware_2012}. Although this basis now only has one free parameter, $D_2$, it still does not correspond to Rossby waves in the linear limit. We can even eliminate this free parameter by interpreting surface buoyancy gradients as topography e.g., by defining 
\begin{equation}
	g_\textrm{buoy} = \left[\frac{f_0^2}{N^2} \d{\psi_B}{z}\right]_{z=z_2}
\end{equation}
where $\psi_B$ corresponds to the background flow, and using $g_\textrm{buoy}$ in place of $g_2$ in the generalized Rhines modes of section \ref{S-linear}\ref{SS-rhines}. However the waves resulting from topographic gradients generally differ from those resulting from vertically-sheared mean-flows (in particular, one must take into account advective continuum modes) and so this resolution is artificial.

\subsection*{Galerkin approximations with $L^2$ modes}
Both the $L^2$ baroclinic modes and the $L^2\oplus \C^2$ modes have infinitely many degrees of freedom. In contrast, numerical simulations only contain a finite number of degrees of freedom. Consequently, it should be possible to use baroclinic modes to produce a Galerkin approximation to quasigeostrophic theory with non-trivial boundary dynamics. Such an approach has been proposed by \cite{rocha_galerkin_2015}.

Projecting $\psi_{\vec k}$ onto the baroclinic modes produces a series expansion, $\psi_{\vec k}^\textrm{exp}$, that is differentiable in the interior but not at the boundaries. By differentiating the series \emph{in the interior} we obtain equation \eqref{qn-trad-expansion} for $q_{\vec k n}$. If instead we integrate by parts twice and avoid differentiating $\psi^\textrm{exp}_{\vec k}$, we obtain
\begin{equation}\label{rocha-qn2}
	q_{\vec k n} = -(k^2+\lambda_n) \psi_{\vec k n} - \frac{1}{H} \sum_{j=1}^2 r_{j\vec k} \, F_n(z_j).
\end{equation}
The two expressions \eqref{qn-trad-expansion} and \eqref{rocha-qn2} are only equivalent when $r_{1\vec k}=r_{2\vec k}=0$. For non-zero $r_{1\vec k}$ and $r_{2\vec k}$, the singular nature of the expansion means we have a choice between equations \eqref{qn-trad-expansion} and \eqref{rocha-qn2}.

By choosing equation \eqref{rocha-qn2} and avoiding the differentiation of $\psi^\textrm{exp}_{\vec k}$, \cite{rocha_galerkin_2015} produced a least-squares approximation to quasigeostrophic dynamics that conserves the surface potential enstrophy integrals \eqref{surface-enstrophy}. This is a conservation property underlying their approximation's success.

\section{Conclusion}
\label{S-conclusion}

In this article, we have studied all possible non-continuum collections of streamfunction normal modes that diagonalize the energy and potential enstrophy. There are four possible $L^2$ modes: the baroclinic modes, the anti-baroclinic modes, the surface modes, and the anti-surface modes. Additionally, we explored the properties of the family of $L^2\oplus\C^2$ bases introduced by \cite{smith_surface-aware_2012} which contain two free parameters $D_1,D_2$ and generalized the family to allow for $D_1,D_2<0$. This generalization is necessary for Rossby waves in the presence of bottom topography. If $D_j=\gamma_j$, where $\gamma_j$ is given by equation \eqref{gamma} for $j=1,2$, the resulting modes are the vertical structure of Rossby waves in a quiescent ocean with prescribed boundary buoyancy gradients (i.e., topography). We have also examined the associated $L^2$ and $L^2\oplus \C^2$ vertical velocity modes.

For the streamfunction $L^2$ modes, only the baroclinic modes are capable of converging pointwise to any quasigeostrophic state on the interval $[z_1,z_2]$, whereas for the vertical velocity $L^2$ modes, only the anti-baroclinic modes are capable. However, in both cases, the resulting eigenfunction expansion is not differentiable at the boundaries, $z=z_1,z_2$. Consequently, while we can recover the volume potential vorticity density, $q_{\vec k}$, we cannot recover the surface potential vorticity densities, $r_{1\vec k}$ and $r_{2 \vec k}$. Thus, we lose two degrees of freedom when projecting onto the baroclinic modes. In contrast, $L^2\oplus \C^2$ modes provide an equivalent representation of the function in question. Namely, the eigenfunction expansion is differentiable on the closed interval $[z_1,z_2]$ so that we can recover $q_{\vec k}$, $r_{1 \vec k}$, $r_{2 \vec k}$ from the series expansion.

We have also introduced a new set of modes, the Rhines modes, that form a basis of $L^2\oplus \C$ and correspond to the vertical structures of Rossby waves over topography. A natural application of these normal modes is to the study of weakly non-linear wave-interaction theories of geostrophic turbulence found in \cite{fu_nonlinear_1980} and \cite{smith_scales_2001}, extending their work to include bottom topography.

\acknowledgments 
We offer sincere thanks to Stephen Garner, Robert Hallberg, Isaac Held, Sonya Legg, and Shafer Smith for comments and suggestions that greatly helped our presentation. 
We also thank Guillaume Lapeyre, William Young, one anonymous reviewer, and the editor (Joseph LaCasce) for their comments that helped us to further refine and focus the presentation, and to correct confusing statements. 
This report was prepared by Houssam Yassin under award NA18OAR4320123 from the National Oceanic and Atmospheric Administration, U.S. Department of Commerce. The statements, findings, conclusions, and recommendations are those of the authors and do not necessarily reflect the views of the National Oceanic and Atmospheric Administration, or the U.S. Department of Commerce.

%
%

%

\appendix[A]\label{A-math}
\appendixtitle{Sturm-Liouville eigenvalue problems with $\lambda$-dependent boundary conditions}

Consider the differential eigenvalue problem
\begin{equation} \label{EigenDiff}
	-\d{}{z}\left(p \, \d{F}{z}\right) + q \, F = \lambda \, r \, F 
\end{equation}
in the interval $(z_1,z_2)$ with boundary conditions
\begin{equation}\label{EigenB}
\begin{split}
	- \left[a_j F - b_j  \left(p  \d{F}{z}\right)(z_j)\right] = \quad \quad \quad \quad \quad \quad \\
		 \lambda \left[c_j  F(z_j) - d_j \left(p \d{F}{z}\right)(z_j)\right]
	\end{split}
\end{equation}
for $j=1,2$, where $1/p(z), q(z), r(z)$ are real-valued integrable functions and $a_j,b_j,c_j,d_j$ are real numbers. Moreover, we assume $p>0, r>0$, that $p$ and $r$ are twice continuously differentiable, that $q$ is continuous, and that $(a_j,b_j)\neq(0,0)$. 

Define the two boundary parameters $D_j$ for $j=1,2$ by
\begin{equation}
	D_j = (-1)^{j+1} \left(a_j\,d_j - b_j\,c_j\right).
\end{equation}
Then the natural inner product for the eigenvalue problem is given by
\begin{equation}\label{appendix-inner}
	\inner{F}{G} = \intz F\,G\,\mathrm{d}z + \sum_{j=1}^{2} D_j^{-1}\left(\mathcal{C}_jF\right) \,\left( \mathcal{C}_j G\right)
\end{equation}
where the boundary operator $\mathcal{C}_j$ is defined by
\begin{equation}
	\mathcal{C}_j F = c_j \, F(z_j) - d_j \,\left( p\,\d{F}{z}\right)(z_j).
\end{equation}

The eigenvalue problem takes place in the space $L^2\oplus \C^N$ where $N$ is the number of non-zero $D_j$. Assume for the following that $N=2$; the case when $N=1$ is similar. If 
\begin{equation}\label{D_j_positive}
	D_j>0
\end{equation}
for $j=1,2$ then the inner product \eqref{appendix-inner} is positive definite\textemdash that is, all non-zero $F$ satisfy $\inner{F}{F}>0$. Therefore $L^2\oplus \C^2$, equipped with the inner product \eqref{appendix-inner}, is a Hilbert space. In this Hilbert space settings, the eigenfunctions $\{F_n\}_{n=0}^\infty$ form and orthonormal basis of $L^2\oplus \C^2$ and that the eigenvalues distinct and bounded below as in equation \eqref{eigen-order} \citep{evans_non-self-adjoint_1970,walter_regular_1973,fulton_two-point_1977}. The appendix of \cite{smith_surface-aware_2012} also proves this result in the case when $d_1=d_2=0$. The convergence properties of normal mode expansions in this case are due to \cite{fulton_two-point_1977}.

However, as we observe in section \ref{S-linear}, the $D_j>0$ case is not sufficient for the Rossby wave problem with topography. In general, the space $L^2\oplus \C^2$ with the indefinite inner product \eqref{appendix-inner} is a Pontryagin space \citep[see][]{iohvidov_spectral_1960,bognar_indefinite_1974}. Pontryagin spaces are analogous to Hilbert spaces except that they have a finite-dimensional subspace of elements satisfying $\inner{F}{F}<0$. If $\Pi$ is a Pontryagin space with inner product $\inner{\cdot}{\cdot}$, then $\Pi$ admits a decomposition 
\begin{equation}
	\Pi = \Pi^{+} \oplus \Pi^{-}
\end{equation}
where $\Pi^{+}$ is a Hilbert space under the inner product $\inner{\cdot}{\cdot}$ and $\Pi^{-}$ is a finite-dimensional Hilbert space under the inner product $-\inner{\cdot}{\cdot}$. If $\{G_n\}_{n=0}$ is an orthonormal basis for the Pontryagin space $\Pi$, then an element $\Psi\in\Pi$ can be expressed
\begin{equation}\label{pont-exp}
	\Psi = \sum_{n=0}^{\infty} \frac{\inner{\Psi}{G_n}}{\inner{G_n}{G_n}}.
\end{equation}
Even though $\{G_n\}_{n=0}^\infty$ is normalized, the presence of $\inner{G_n}{G_n}=\pm 1$ in the denominator of equation \eqref{pont-exp} is essential since this term may be negative.

One can rewrite the eigenvalue problem \eqref{EigenDiff}\textendash\eqref{EigenB} in the form $\mathcal{L}\, F = \lambda \, F$ for some operator $\mathcal{L}$ \cite[e.g.,][]{langer_spectral_1991}. The operator $\mathcal{L}$ is a positive operator if 
\begin{itemize}
	\item  for the $\lambda$-dependent boundary conditions, we have 
			\begin{equation}\label{left-definite-1}
				\frac{a_i \, c_i}{D_i} \leq 0, \quad \frac{b_i \, d_i}{D_i} \leq 0, \quad (-1)^i \frac{a_i \, d_i}{D_i} \geq 0
			\end{equation} 
	\item for the $\lambda$-independent boundary conditions, we have 
			\begin{equation}\label{left-definite-2}
				b_i = 0  \quad \text{or} \quad  (-1)^{i+1}\frac{a_i}{b_i} \geq 0 \quad \text{ if } b_i \neq 0.
			\end{equation}
\end{itemize}
\cite{yassin_normal_nodate} has shown that, when $\mathcal{L}$ is positive, the eigenfunctions $\{F_n\}_{n=0}^\infty$ of the eigenvalue problem \eqref{EigenDiff}\textendash\eqref{EigenB} form an orthonormal basis of $L^2\oplus{\C}^2$, that the eigenvalues are real, and that the eigenvalues are ordered as in equation \eqref{eigen-order}. Moreover, since $\mathcal{L}$ is positive, we have the relationship
\begin{equation}
	\lambda \inner{F}{F} = \inner{\mathcal{L}F}{F} \geq 0.
\end{equation}
Finally, \cite{yassin_normal_nodate} shows that the normal mode expansion results of \cite{fulton_two-point_1977} extend to this case as well.
 
\appendix[B]
\appendixtitle{Polarization relations and the vertical velocity eigenvalue problem}

\subsection{Polarization relations}
 

The linear quasigeostrophic vorticity and buoyancy equations, computed about a resting background state, are
\begin{align}
	\label{zeta-equation-linear}
	\pd{\zeta}{t} + \beta \, \pd{\psi}{x} &= f_0 \pd{w}{z}, \\
	\label{b-equation-linear}
	\pd{b}{t} &= -N^2 \, w
\end{align}
in the interior $z\in(z_1,z_2)$. The vorticity, $\zeta$, and buoyancy, $b$, are given in terms of the geostrophic streamfunction via 
\begin{align}
	\label{zeta}
	\zeta = \lap \psi \\
	\label{b}
	b = f_0 \pd{\psi}{z},
\end{align} 
The no-normal flow at the lower and upper boundaries implies
\begin{equation}\label{vv-boundary}
	f_0 \, w = \vec u \cdot \grad g_j
\end{equation}
for $j=1,2$. Substituting equation \eqref{vv-boundary} into the linear buoyancy equation \eqref{b-equation-linear}, yields the boundary conditions
\begin{equation}\label{buoyancy-boundary-evolution}
	\partial_t b + \vec u \cdot \grad \left(\frac{N^2}{f_0} g_j\right) = 0 \quad \textrm{for } z=z_j.
\end{equation}

We now assume solutions of the form 
\begin{equation}
	\psi= \hat \psi(z) \, e_{\vec k}(\vec x) \, \mathrm{e}^{-\mathrm{i}\omega t}
\end{equation}
and similarly for $w$. Substituting such solutions into equations \eqref{zeta-equation-linear}\textendash\eqref{b-equation-linear} and using $\vec u = \unit z \times \grad \psi$ gives
\begin{align}
	\label{psiz-w}
	\d{\hat \psi}{z} &= -\mathrm{i} \, \frac{N^2}{f_0\, \omega} \hat w\\
	\label{wz-psi}
	\d{\hat w}{z} &= \mathrm{i} \, \frac{\omega}{f_0} \left[k^2 + \frac{\beta \, k_x}{\omega} \right] \hat \psi.
\end{align}
for $z \in (z_1,z_2)$. 
At the boundaries $z=z_1,z_2$, we use equations \eqref{vv-boundary} and \eqref{buoyancy-boundary-evolution} to obtain
\begin{align}
	\label{bhat-boundary}
	\hat b &= - \frac{N^2}{f_0 \, \omega} \, \hat{\vec u} \cdot \grad g_j \\
	\label{what-boundary}
	\hat w &= \mathrm{i} \frac{1}{f_0} \, \hat{\vec u} \cdot \grad g_j.
\end{align}

\subsection{The vertical velocity eigenvalue problem}

Taking the vertical derivative of \eqref{wz-psi} and using \eqref{psiz-w} yields
\begin{equation}\label{vertical-velocity-interior}
	- \dd{\chi}{z} = \lambda \left(\frac{N^2}{f_0^2}\right)  \chi 
\end{equation}
where $\hat w = w_0 \, \chi(z)$ and $\chi$ is non-dimensional. The boundary conditions at $z = z_{j}$ are
\begin{equation}\label{vertical-velocity-boundary-again}
	- (-1)^j \, \gamma_j \, k^2 \, \d{\chi}{z} = \lambda \left[\chi + (-1)^j \, \gamma_j \, \d{\chi}{z} \right]
\end{equation}
as obtained by using equations \eqref{wz-psi} and \eqref{psiz-w} in boundary conditions \eqref{Rhines-eigen-boundary}. The orthonormality condition is
\begin{align}\label{vertical-velocity-ortho}
\begin{split}
	\pm \delta_{mn} = \frac{1}{H} &\left[ \intz \chi_m \, \chi_n \left( \frac{N^2}{f_0^2}\right) \mathrm{dz} \right. \\ & \left.  -\frac{1}{k^2} \sum_{j=1}^2 \frac{1}{\gamma_j} \left(\mathcal{C}_j \chi_m \right) \left(\mathcal{C}_j \chi_n\right)  \right] 
	\end{split}
\end{align}
where 
\begin{equation}
	\mathcal{C}_j \chi = \chi(z_j) + (-1)^j \, \gamma_j \d{\chi(z_j)}{z}.
\end{equation}

When only one boundary condition is $\lambda$-dependent (e.g., $\gamma_2=0$) the eigenvalue problem \eqref{vertical-velocity-interior}\textendash\eqref{vertical-velocity-boundary-again} satisfies equation \eqref{D_j_positive} when $\gamma_1>0$ and equations \eqref{left-definite-1} and \eqref{left-definite-2} when $\gamma_1<0$; thus the reality of the eigenvalues and the completeness results follow. However, when both boundary conditions are $\lambda$-dependent the problem no longer satisfies these conditions for all $\vec k$. Instead, in this case, one exploits the relationship between the vertical velocity eigenvalue problem \eqref{vertical-velocity-interior}\textendash\eqref{vertical-velocity-boundary-again} and the streamfunction problem \eqref{Rhines-eigen-interior}\textendash\eqref{Rhines-eigen-boundary} given by equations \eqref{psiz-w} and \eqref{wz-psi} to conclude that the two problem have the identical eigenvalues (for $\omega \neq 0$) and then use the simplicity of the eigenvalues to conclude that no generalized eigenfunctions can arise.

\subsection{The vertical velocity $L^2$  modes} 
 
Analogously with the streamfunction $L^2$ modes, we have the following sets of vertical velocity $L^2$ modes.

\begin{itemize}
	\item \textit{Baroclinic modes}: Vanishing vertical velocity at both boundaries,
	\begin{equation}
		\chi(z_1) = 0, \quad \chi(z_2) = 0.
	\end{equation}
	\item \textit{Anti-baroclinic modes}: Vanishing pressure at both boundaries,
	\begin{equation}
		\d{\chi(z_1)}{z} = 0, \quad \d{\chi(z_2)}{z}=0.
	\end{equation}
	\item \textit{Surface modes}:
	\begin{equation}
		\d{\chi(z_1)}{z} = 0, \quad \chi(z_2) = 0.
	\end{equation}
	\item \textit{Anti-surface modes}:
	\begin{equation}
		\chi(z_1) =0, \quad \d{\chi(z_2)}{z} = 0. 
	\end{equation} 
\end{itemize}






%
%
%
\bibliographystyle{ametsoc2014}
\bibliography{references}

%

%

\end{document}